\documentclass[manuscript,screen]{acmart}

\usepackage{graphicx}%
\usepackage{multirow}%
\usepackage{amsmath}

\usepackage{amssymb,amsfonts}
\usepackage{amsthm}%
\usepackage{mathrsfs}%
\usepackage[title]{appendix}%
\usepackage{xcolor}%
\usepackage[table]{xcolor}
\usepackage{textcomp}%
\usepackage{manyfoot}%
\usepackage{booktabs}%
\usepackage{algorithm}%
\usepackage{algorithmicx}%
\usepackage{algpseudocode}%
\usepackage{listings}%
\usepackage{caption}
\usepackage{makecell}
\usepackage{longtable}
\usepackage{tabularx}
\usepackage[most]{tcolorbox} 
\usepackage{array}
\usepackage{placeins}
\AtBeginDocument{%
  }

\begin{document}

\title[NL–PL Software Traceability Link Recovery Needs More than Textual Similarity]{Natural Language–Programming Language Software Traceability Link Recovery Needs More than Textual Similarity}

\author{Zhiyuan Zou}
\email{Zzy2499488704@163.com}
\affiliation{%
  \institution{School of Computer Science and Artificial Intelligence, Wuhan Textile University}
  \city{Wuhan}
  \state{Hubei}
  \country{China}
}

\author{Bangchao Wang}
\email{wangbc@whu.edu.cn}
\authornotemark[1]
\affiliation{%
  \institution{School of Computer Science and Artificial Intelligence, Wuhan Textile University}
  \city{Wuhan}
  \state{Hubei}
  \country{China}}

\author{Peng Liang}
\email{liangp@whu.edu.cn}
\affiliation{%
  \institution{School of Computer Science, Wuhan University}
  \city{Wuhan}
  \state{Hubei}
  \country{China}
}

\author{Tingting Bi}
\email{Tingting.Bi@unimelb.edu.au}
\affiliation{
\institution{The University of Melbourne and The University of Western Australia}
\country{Australia}
}

\author{Huan Jin}
\email{jh\_0230@163.com}
\affiliation{%
 \institution{School of Computer Science and Artificial Intelligence, Wuhan Textile University}
  \city{Wuhan}
  \state{Hubei}
  \country{China}
}

\renewcommand{\shortauthors}{Zou et al.}

\begin{abstract}
In the field of software traceability link recovery (TLR), textual similarity has long been regarded as the core criterion. However, in tasks involving natural language and programming language (NL-PL) artifacts, relying solely on textual similarity is limited by their semantic gap. To this end, we conducted a large-scale empirical evaluation across various types of TLR tasks, revealing the limitations of textual similarity in NL-PL scenarios. To address these limitations, we propose an approach that incorporates multiple domain-specific auxiliary strategies, identified through empirical analysis, into two models: the Heterogeneous Graph Transformer (HGT) via edge types and the prompt-based Gemini 2.5 Pro via additional input information.
We then evaluated our approach using the widely studied requirements-to-code TLR task, a representative case of NL-PL TLR. Experimental results show that both the multi-strategy HGT and Gemini 2.5 Pro models outperformed their original counterparts without strategy integration. Furthermore, compared to the current state-of-the-art method HGNNLink, the multi-strategy HGT and Gemini 2.5 Pro models achieved average F1-score improvements of 3.68\% and 8.84\%, respectively, across twelve open-source projects, demonstrating the effectiveness of multi-strategy integration in enhancing overall model performance for the requirements-code TLR task.
\end{abstract}



\maketitle

\section{Introduction}\label{introduction}
Software Traceability (ST) refers to the ability to describe and track the lifecycle of software artifacts, both in forward and backward directions \cite{gotel1994analysis, pinheiro2004requirements}. Maintaining traceability between software artifacts is essential throughout the software development lifecycle. As Watkins et al. \cite{watkins1994and} once stated, ``\textit{You can't manage what you can't trace}''. ST plays a critical role in various software engineering tasks such as change impact analysis, safety analysis, and coverage analysis \cite{cleland2014software, rath2018traceability}. Among its many aspects, Traceability Link Recovery (TLR) is a key research area within ST, aiming to automatically establish trace links between different software artifacts \cite{antoniol2002recovering, van2023effectiveness}. However, manually creating such traceability links is often labor-intensive and time-consuming \cite{mahmoud2012semantic, lin2021traceability, egyed2010effort, rath2018traceability}.

Therefore, researchers have proposed various automated traceability link recovery methods, among which information retrieval (IR)-based, learning-based, pre-trained based, and heuristic-based approaches have been extensively studied and applied \cite{aung2020literature, wang2024empirical}. Among these methods, textual similarity plays a central role - either as a direct decision criterion or as a feature for learning-based models.

Existing text similarity-based methods exhibit a significant performance gap between natural language-to-natural language (NL-NL) or programming language-to-programming language (PL-PL) artifact pairs and natural language-to-programming language (NL-PL) artifact pairs. For example, in the Issue-Commit task (an NL-NL TLR task), the current state-of-the-art (SOTA) method, MPLinker \cite{wang2025mplinker}, constructs prompt templates using the textual descriptions of Issues and Commits, and then utilizes a large language model (LLM) to determine the presence of links based on their textual similarity. MPLinker achieves an average F1-score of approximately 0.95 across multiple projects, demonstrating outstanding performance. In contrast, in the Requirements-Code task (an NL-PL TLR task), the current SOTA method HGNNLink \cite{wang2025HGNNLink} encodes both requirement artifacts and code artifacts using pre-trained models. A heterogeneous graph neural network (HGNN) then determines whether a link (i.e., a traceability relationship) exists between a requirement and a code artifact based on the similarity of their encoded representations. However, HGNNLink only achieves an average F1-score of around 0.70 across multiple projects - showing a notable performance gap compared to MPLinker.

To further investigate the causes of the aforementioned phenomenon, we analyzed several projects involving NL-PL TLR tasks and identified the following key factors. First, PL artifacts are generally more abstract in nature and, aside from minimal comments, often lack clear descriptions of functional semantics. Their content is primarily composed of keywords and complex API calls, many of which are semantically unrelated to the detailed functionalities described in corresponding NL artifacts. This leads to a semantic mismatch between the two artifact types. Second, due to the modular nature of PL artifacts, a single NL artifact often corresponds to multiple PL artifacts. Some of these code segments implement functionalities that are only briefly mentioned in the corresponding NL artifacts, resulting in low textual similarity, even though valid traceability links do exist in actual development scenarios. Second, due to the modular nature of PL artifacts, a single NL artifact often corresponds to multiple PL artifacts. Some of these code segments implement functionalities that are only briefly mentioned in the corresponding NL artifacts, resulting in low textual similarity, even though valid traceability links do exist in actual development scenarios. For example, in the Requirements-to-Code TLR task from the iTrust project\footnote{http://sarec.nd.edu/coest/datasets.html}, the NL artifact \textsc{UC3.txt} describes a user authentication use case, while the PL artifact \textsc{EmailUtil.java} implements functionalities related to email sending. Although the requirement document does mention email functionality, it appears only once as a minor step in a sub-flow. Furthermore, \textsc{EmailUtil.java} contains a large number of API calls related to database operations, which are not textually related to the requirement description in any way. In contrast, for NL-NL TLR tasks, since both artifacts are expressed in natural language and lack the structured nature of code, the textual similarity between linked artifacts tends to be much more prominent.

The above observations raise a critical question: \textbf{Is relying solely on textual similarity sufficient for NL-PL TLR tasks, or should richer strategies be introduced to achieve more accurate and robust traceability link recovery?}

\begin{figure}
    \centering
    \includegraphics[width=\linewidth]{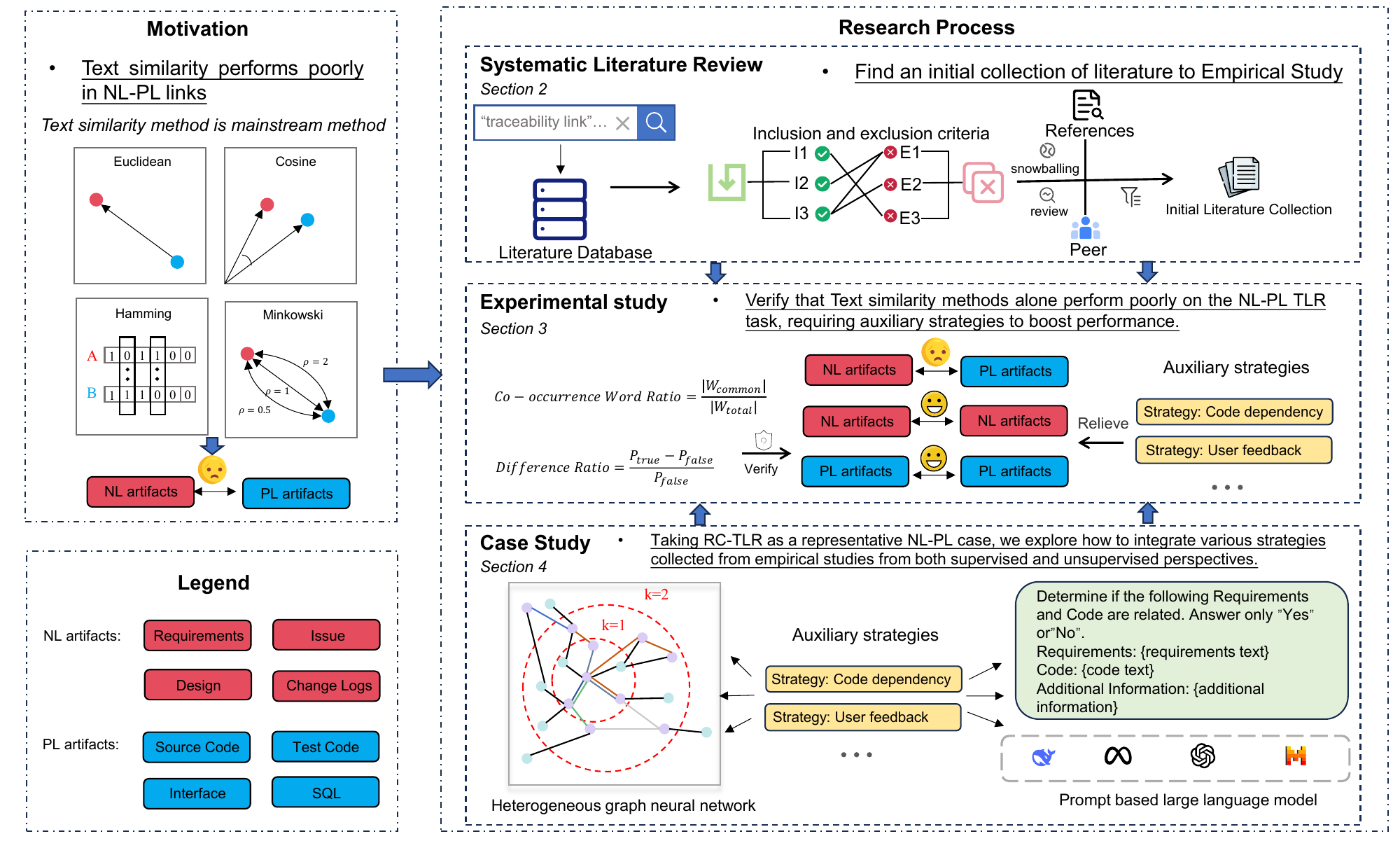}
    \caption{Research process with the motivation}
    \label{research process}
\end{figure}

To address this question, this study first conducted a systematic review of research in this field over the past five years. We performed a statistical analysis of the projects and strategies used in existing work and evaluated the effectiveness of textual similarity-based methods across different types of projects. The findings reveal that for NL-PL TLR tasks, the performance of purely text-based approaches is significantly limited. Meanwhile, a growing body of research has attempted to introduce additional strategies to improve link quality. Building on these insights, we selected the Requirements-to-Code TLR task as a representative scenario of NL-PL TLR and proposed two strategy-integrated models tailored to supervised and unsupervised settings. The first method is based on the Heterogeneous Graph Transformer (HGT) \cite{hu2020heterogeneous}, while the second leverages the large-scale pre-trained language model Gemini 2.5 Pro \cite{team2023gemini} with a prompt-learning mechanism. Both approaches demonstrate substantial improvements in TLR performance for the Requirements-to-Code task. The detailed research motivation and process are illustrated in Figure \ref{research process}. The \textbf{main contributions} are as follows:
\begin{enumerate}
    \item We first systematically organized and classified 99 related projects and their applied strategies in the software traceability domain over the past five years, released the results publicly, and constructed a standardized benchmark to support future research.
    \item On this basis, we proposed a novel low-cost evaluation metric for text similarity, namely the Difference Ratio, and conducted experiments on the collected 99 projects. The empirical results demonstrate that methods relying solely on textual similarity perform poorly in NL-PL TLR tasks.
    \item Based on this insight, we proposed a supervised approach built upon a Heterogeneous Graph Transformer, which integrates multiple auxiliary strategies (e.g., code dependencies, user feedback, fine-grained text similarity) to enhance model performance.
    \item Furthermore, in the unsupervised setting, we introduced a prompt-learning mechanism that embeds multiple auxiliary strategies as additional information into prompt templates, and applied it to the Gemini 2.5 Pro model, thereby developing an efficient TLR solution that requires no labeled data.
\end{enumerate}

The rest of this paper is organized as follows: Section \ref{slr} outlines the process of the systematic literature review. Section \ref{experimental_study} presents an experimental study on the collected literature in terms of projects and strategies. Section \ref{sec case study} proposes and describes two types of methods through a case study. Section \ref{sec discussion} discusses the interpretation of RQs and implications for researchers and practitioners. Section \ref{tv} discusses the threats to the validity of this study. Section \ref{related work} discusses the related work of this study, Section \ref{conclusion} concludes the work and outlines the directions for future work.

\section{Systematic Literature Review}\label{slr}
\begin{figure}
    \centering
    \includegraphics[width=\linewidth]{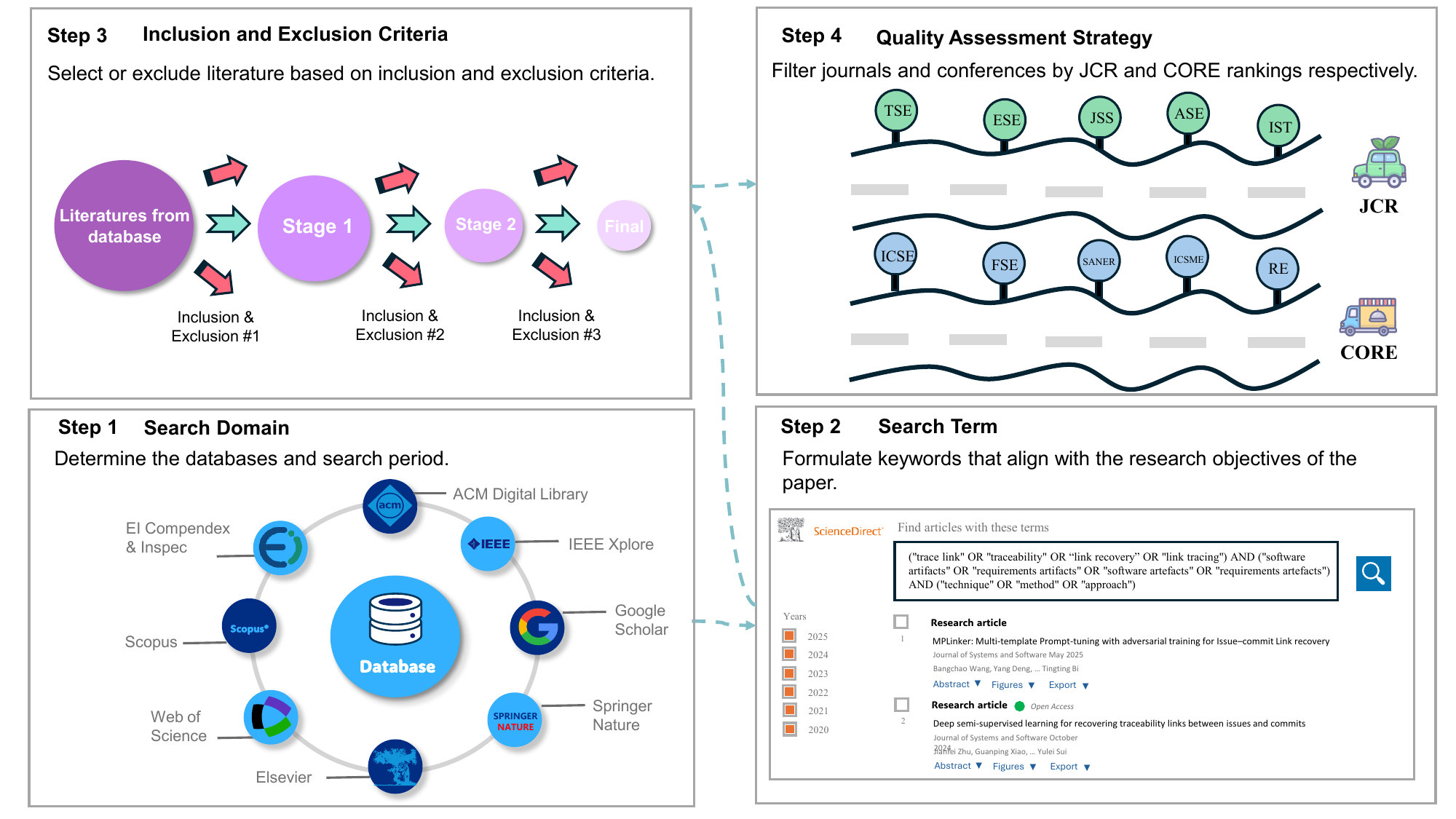}
    \caption{Overview of the literature search process}
    \label{initial_set_of_literature_search_process}
\end{figure}

In this section, we discuss how this study searches for relevant literature in the field of software traceability and presents the search results. To ensure the high relevance and fairness of the set of research literature, we elaborate on the search methodology based on the recommendations of Petersen et al. \cite{petersen2015guidelines}, covering aspects such as search domains, search terms, inclusion and exclusion criteria, and quality assessment strategies. The literature search process is shown in Figure \ref{initial_set_of_literature_search_process}. Finally, we present the results after the literature search.

\subsection{Search Domain}\label{search domain}
According to the recommendations of \cite{wang2024empirical}, \cite{olivero2024systematic}, and \cite{pauzi2023applications}, we selected eight literature databases for the literature search, including \textit{ACM Digital Library\footnote{https://dl.acm.org/}}, \textit{EI Compendex \& Inspec\footnote{https://www.engineeringvillage.com/}}, \textit{IEEE Xplore\footnote{https://ieeexplore.ieee.org/Xplore/home.jsp}}, \textit{Science Direct\footnote{https://www.sciencedirect.com/}}, \textit{SpringerLink\footnote{https://link.springer.com/}}, \textit{Web of Science\footnote{https://www.webofscience.com/wos/woscc/basic-search}}, \textit{Scopus\footnote{https://www.scopus.com/home.uri}}, and \textit{Google Scholar\footnote{https://scholar.google.com/}}.

The primary purpose of conducting a systematic literature review in this study is to obtain a set of research literature from which data can be extracted to verify the motivations mentioned in Section \ref{introduction}. To explore the current state of research in this field while considering the timeliness of the literature, this study only includes relevant studies from the period between 2020 and 2024, as well as studies from January 1, 2025, to June 1, 2025.

\subsection{Search Terms}\label{search terms}
To ensure the comprehensiveness and relevance of the search results, this study formulated the search terms based on the PICO framework \cite{petersen2015guidelines}.

\textbf{Population:} The ``population'' in this study is ``traceability''. We also used synonyms of ``traceability'' in the field, such as ``trace link'' and ``link recovery'', to expand the scope of the search \cite{wang2018requirements}.

\textbf{Intervention:} The ``Intervention'' in this study consists of both ``Domain Intervention'' and ``Paper Type Intervention''. We used ``software artifacts'' to define the scope of the tracking task in this study. Furthermore, since requirements traceability is an important research direction in software traceability, and experiments have shown that including ``requirements artifacts'' in the search terms significantly expands the literature search scope, we also included ``requirements artifacts'' as a ``Domain Intervention'' term. Additionally, based on the suggestion from Pauzi et al. \cite{pauzi2023applications}, since the term ``artifacts'' is spelled differently in British and American English, we included ``software artifacts'' and ``requirements artifacts'' as search terms. For ``Paper Type Intervention'', this study only considers innovative research and excludes review-based research. Therefore, we included ``technique'', ``method'', and ``approach'' as the ``paper type intervention'' terms.

\textbf{Comparison:} This study focuses on the determination methods and artifact categories for link generation in the initial studies, without considering which other studies they have been or need to be compared with. Therefore, the ``Comparison'' aspect is not relevant to this study.

\textbf{Outcome:} The data in this study are derived from specific experiments and do not rely on the conclusions of the initial literature. Therefore, the ``Outcome'' aspect is not relevant to this study. 

The search keywords for the aspects of ``Population'', ``Domain Intervention'', and ``Paper Type Intervention'' aspects are connected using \textbf{``AND''}, while the keywords within each aspect are connected using \textbf{``OR''}. The search keywords for each aspect are shown in Table \ref{Literature search terms}.

\begin{table}[ht]
\centering
\caption{Literature search terms.}
\label{Literature search terms}
\begin{tabular}{@{}p{0.25\textwidth} p{0.75\textwidth}@{}}
\toprule
\textbf{PICO} & \textbf{Search terms} \\ \midrule
Population & trace link, traceability, link recovery \\
Domain Intervention & software artifacts, requirements artifacts, software artefacts, requirements artefacts\\
Paper Type Intervention & technique, method, approach\\
\bottomrule
\end{tabular}
\end{table}

\subsection{Inclusion and Exclusion Criteria}\label{in_exclusion}
To exclude irrelevant literature and ensure the quality of the initial literature collection, we established a set of inclusion and exclusion criteria, as shown in Table \ref{Inclusion and exclusion criteria}, based on the studies by Wang et al. \cite{wang2024empirical}, Mucha et al. \cite{mucha2024systematic} and Hassan et al. \cite{hassan2024systematic}. To enhance the completeness of the initial literature, we performed snowballing \cite{wohlin2014guidelines} on the remaining literature after applying the inclusion and exclusion criteria. This process involved both forward and backward snowballing to further include other relevant studies.

\begin{table}[ht]
\centering
\caption{Inclusion and exclusion criteria.}
\label{Inclusion and exclusion criteria}
\begin{tabular}{@{}p{0.1\textwidth} p{0.9\textwidth}@{}}
\toprule
\textbf{No.} & \textbf{Selection criteria} \\ \midrule
I1 & Select literature written in English. \\
I2 & Select relevant literature in the field of software traceability.\\
\midrule
E1 & Exclude editorials, white papers, catalogs, extend abstracts, communications, books, tutorials, PowerPoint and non-peer-reviewed papers.\\
E2 & Exclude literature less than six pages (excluding six pages).\\
E3 & Exclude gray literature.\\
E4 & Exclude duplicate literature.\\ 
E5 & Exclude review literature.\\
E6 & Exclude literature that does not evaluate the method(s).\\
\bottomrule
\end{tabular}
\end{table}

\subsection{Quality Assessment Strategy}\label{quality}
To ensure the quality of the selected articles, based on the suggestions of \cite{mosquera2024understanding} and \cite{ouhbi2015requirements}, we use the Journal Citation Reports\footnote{https://www.scimagojr.com} (JCR) to rate the journal articles. Specifically, if the journal's ranking is Q2 or higher, the article will be included in the initial literature set; otherwise, it will be excluded. For conference papers, we refer to the CORE ranking\footnote{https://portal.core.edu.au/conf-ranks/}, and if the conference ranking is C or higher, the corresponding paper will be selected for the initial literature set; otherwise, it will also be excluded.

\subsection{Search Results}\label{sec3.5}
\begin{figure}
    \centering
    \includegraphics[width=\linewidth]{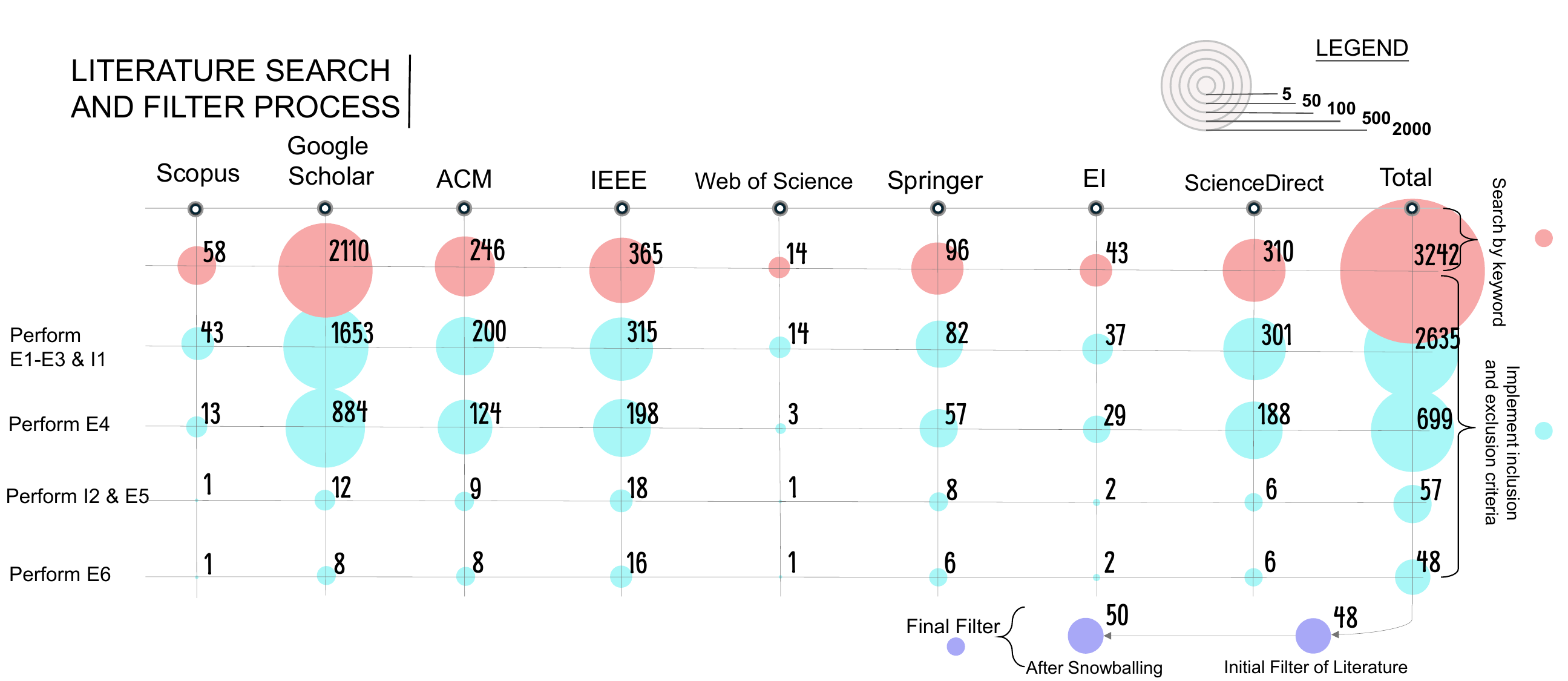}
    \caption{Literature search and filtering process}
    \label{search_filtering}
\end{figure}

Based on the literature search methods described in Sections \ref{search domain} to \ref{quality}, we performed searches in eight databases of the literature and screened the search results. The literature screening process is shown in Figure \ref{search_filtering}. We initially performed a search in eight databases of literature based on keywords and publication year restrictions, obtaining a preliminary set of 3,242 relevant articles. Subsequently, we performed the first screening round using the exclusion criteria E1–E3 and the inclusion criteria I1, combined with the metadata of the articles, leaving 2,635 articles. Next, we used the Zotero tool to remove duplicates (applying the exclusion criterion E4), retaining 699 articles after screening. Based on this, we further reviewed the abstracts, introductions, and methodology sections of the articles, applying the inclusion criterion I2 and the exclusion criterion E5 for screening, ultimately retaining 57 articles. We then evaluated the results sections of these articles (applying the exclusion criterion E6), excluding an additional 9 articles, leaving 48. Finally, through snowballing supplementary searches, we identified 2 eligible articles that were not covered in the initial search. Therefore, this study ultimately included 50 articles as the initial literature set. Detailed descriptive information for each article can be found in the supplementary materials \cite{source_code}.

\section{Experimental Study}\label{experimental_study}
In this section, we conducted a statistical analysis of the literature data collected in Section \ref{sec3.5}. Through a series of empirical studies, we evaluate whether textual similarity is sufficient to support the needs of traceability link recovery tasks in different types of projects. In addition, we investigate the alternative methods employed by researchers to compensate for the limitations of textual similarity.

\subsection{Evaluation Metrics}\label{sec_metric}
In the requirements-code traceability link recovery task, Precision, Recall, and F1-score are the three most commonly used evaluation metrics \cite{wang2024empirical, wang2022systematic}. These metrics effectively measure a model’s ability to identify correct traceability links, helping to assess its accuracy and completeness in practical traceability tasks.

Precision indicates the proportion of correctly identified trace links among all the links predicted as positive by the model. It is defined as:
\begin{equation}
    Precision = \frac{Ture\,Positives\,(TP)}{True\,Positives\,(TP)\,+\, False \, Positives\,(FP)}
\end{equation}

Recall measures the proportion of actual positive links that are correctly identified by the model. It is defined as:
\begin{equation}
    Recall = \frac{True\,Positives\,(TP)}{True\,Positives\,(TP)\,+\,False\,Negatives\,(FN)}
\end{equation}

F1-score is the harmonic mean of Precision and Recall, providing a balanced evaluation between accuracy and completeness. It is defined as:
\begin{equation}
    F1-Score = \frac{2*Precision*Recall}{Precision+Recall}
\end{equation}

\subsection{Research Questions}\label{research_question}
To investigate whether textual similarity methods are sufficient to meet the needs of NL-PL software artifact traceability tasks, this section proposes three research questions (RQs).
RQ1 aims to explore existing TLR projects that involve different types of artifacts and to provide experimental subjects for the subsequent experiments in this study.
RQ2 seeks to determine whether using only textual similarity methods can satisfy the requirements of NL-PL software artifact traceability tasks.
RQ3 focuses on identifying the auxiliary strategies adopted by researchers in this field to address the shortcomings of relying solely on textual similarity methods.
The rationale behind each RQ and the experimental details are presented in the following subsections.

\subsubsection{RQ1: Which projects are commonly used as research subjects in the field of software traceability?}

To select appropriate subjects for the subsequent experiments, we first propose \textit{RQ1: Which projects are commonly used as research subjects in the field of software traceability?}

To answer this RQ, we extracted the projects used in the 50 papers collected in Section \ref{sec3.5} and perform a statistical analysis from the following five perspectives: the number of source artifacts, the number of target artifacts, the number of true links, the types of source-target artifact pairs, and the reference for the project.

\subsubsection{RQ2: How effective is traceability link generation based on textual similarity across different types of project artifacts?}
Subsequently, we analyze the difficulty of performing traceability link prediction using textual similarity across different projects by comparing the proportion of shared terms between artifacts in true links and those in non-true links. Accordingly, we propose \textit{RQ2: How effective is traceability link generation based on textual similarity across different types of project artifacts?}

To further evaluate the effectiveness and difficulty of traceability link recovery tasks based on textual similarity in different projects, we design a metric: \textbf{the difference in co-occurrence word ratios between true and non-true sets (hereafter referred to as the ``Difference Ratio'')}. This metric is derived based on the statistics of artifact pairs in true and false links.

Specifically, we first calculate the co-occurrence word ratio for artifact pairs in the true and non-true sets:
\begin{equation}
    Co-occurrence\, Word\, Ratio = \frac{|W_{common}|}{|W_{total}|} 
\end{equation}
where $W_{common}$ represents the set of common words between the two artifacts (i.e., words that appear in both artifacts), and $W_{total}$ represents the union of all distinct words from both artifacts. This ratio measures the lexical similarity between two artifacts.

Next, based on the difference in average co-occurrence word ratios between true and false links, we calculate the ``Difference Ratio'' metric, as shown below:
\begin{equation}
    Difference \, Ratio = \frac{P_{true}-P_{false}}{P_{false}}
\end{equation}
where $P_{true}$ represents the average co-occurrence word ratio for true link artifact pairs, and $P_{false}$ represents the average co-occurrence word ratio for false link artifact pairs. A larger difference degree indicates that textual similarity is more effective at distinguishing true and false links in the project, suggesting that the traceability link recovery task is easier and more effective based on textual similarity.

Given that true and false links are typically imbalanced in number, we mitigate this bias by randomly sampling from the false link set, ensuring the sample size matches that of the true link set. To reduce the variance introduced by random sampling, we perform the sampling 100 times, calculating the co-occurrence word ratio for each sample and taking the average as the final $P_{false}$.

Additionally, to test whether the difference in co-occurrence word ratios between true and false links is statistically significant, we use a non-parametric hypothesis testing method—Mann-Whitney U test. To aggregate the results of multiple samples, we apply the Fisher method to combine the p-values from 100 independent tests. Specifically, we first perform the Mann-Whitney U test between the true link set and each randomly sampled subset of false links (with the same size as the true link set) to obtain 100 independent p-values. Then, we calculate the combined \textit{p-value} using the Fisher statistic:
\begin{equation}
    \chi=-2\sum ln(p_i)
\end{equation}

where the Fisher statistic follows a chi-squared distribution with 200 degrees of freedom.

The $H_0$ is that the co-occurrence word ratio in true links is not significantly higher than that in false links, while the $H_1$ is that the co-occurrence word ratio in true links is significantly higher than in false links.

The Mann-Whitney U value calculation is as follows:
\begin{equation}
    U=N_AN_B+\frac{N_A(N_A+1)}{2}-R_1
\end{equation}
\begin{equation}
    U'=N_AN_B-U
\end{equation}

where $N_A$ and $N_B$ represent the sample sizes of the two experimental groups (true set and false set) with $N_A\leq N_B$, and $R_1$ is the sum of ranks for the true set. For a single test, if $min(U,U')$ is less than the critical value corresponding to the significance level $\alpha<0.01$, the $H_0$ is rejected. For the combined test, if the \textbf{p-value} obtained through the Fisher method is less than 0.01, the null hypothesis is rejected, indicating that the co-occurrence word ratio in true links is significantly higher than in false links.

In addition, to further verify whether the proposed “Difference Ratio” metric is statistically significantly associated with the effectiveness of text similarity-based methods in TLR tasks, we conducted an experimental study using the RoBERTa model across all projects. IR-based methods and pretrained language model-based methods are the most commonly used text similarity-based approaches in the TLR domain. However, when applied to large-scale datasets, IR methods often result in highly sparse similarity matrices, leading to substantial computational overhead. Therefore, considering experimental efficiency and computational resource constraints, we did not adopt IR methods in this study. Instead, we selected RoBERTa due to its strong generalization capabilities and its demonstrated performance in TLR tasks, as shown in prior studies by \cite{lan2023btlink} and \cite{zhu2024deep}. To limit computational costs, we use RoBERTa as the sole representative model to examine the correlation between the “Difference Ratio” metric and the effectiveness of text similarity-based TLR methods.

For the experimental design, we grouped the projects based on their respective categories and calculated the average F1 performance of the RoBERTa model on TLR tasks within each group. Each project group was treated as an observation sample, and we used Spearman’s rank correlation coefficient to analyze the relationship between each group’s “Difference Ratio” and its average F1-score under RoBERTa. Spearman’s rank correlation is a non-parametric test suitable for small sample sizes and data that do not follow a normal distribution, aligning well with the context of this study.

The Spearman correlation coefficient $\rho$ is calculated as follows:
\begin{equation}
    \rho = 1- \frac{6\sum d_i^2}{n(n^2-1)}
\end{equation}

where $d_i$ denotes the difference in ranks for the $i$-th project between the two variables (“Difference Ratio” and F1-score), and $n$ is the number of project groups.

Since the Spearman coefficient $\rho$ reflects only the strength and direction of a monotonic relationship, we further calculate its corresponding \textit{p-value} to determine the statistical significance of the observed correlation.

To compute the \textit{p-value}, we employed an exact permutation test. This method fixes the ranks of the “Difference Ratio” values and randomly permutes the ranks of the average F1 values across different groups to generate a large number of random permutations. For each permutation, a new Spearman correlation coefficient $\rho_{perm}$ is computed. Then, we calculate the two-tailed \textit{p-value} using the following formula:

\begin{equation}
    p=\frac{Number\, of \, |\rho_{perm}|\geq|\rho|}{Total\, permutations}
\end{equation}

The resulting \textit{p-value} is then used to test the following hypotheses:
\begin{itemize}
    \item $H_2$: The “Difference Ratio” metric is not significantly correlated with the effectiveness of text similarity-based methods in TLR tasks.
    \item $H_3$: The “Difference Ratio” metric is significantly correlated with the effectiveness of text similarity-based methods in TLR tasks
\end{itemize}

If the \textit{p-value} is less than the significance level $\alpha=0.05$, we reject the null hypothesis $H_2$, indicating that the correlation between the two variables is statistically significant.

\subsubsection{RQ3: What strategies can support text similarity-based methods in generating more accurate trace links for different types of artifacts?}

To explore the strategies proposed in current research that enhance the effectiveness of text similarity-based methods, we formulated \textit{RQ3: What strategies can support text similarity-based methods in generating more accurate trace links for different types of artifacts?}

To answer this question, we conducted a statistical analysis of the strategies discussed in the 50 key studies collected in Section \ref{sec3.5}. We not only categorized and summarized the strategies but also provided concrete examples to help researchers better understand how these strategies are applied in practice.

In addition, we further integrated the findings of RQ2 to investigate the relationship between the difficulty of traceability link recovery across different artifact combinations and the number of enhancement strategies proposed for each. Through this correlation analysis, our aim was to determine whether researchers have introduced more targeted strategies for heterogeneous artifact combinations - those with lower differential ratio values - to compensate for the inherent limitations of text similarity-based methods in such contexts.

\subsection{Results}

\subsubsection{Answer RQ1: Which projects are commonly used as research subjects in the field of software traceability? }
The experimental results of RQ1 are presented in Table \ref{table app A} of Appendix A. From the 50 primary studies selected in Section \ref{sec3.5}, we extracted a total of 99 open-source projects, covering nine types of traceability artifact pairs: Issue-Commit, Issue-Issue, Requirements-Code, Requirements-Requirements, Requirements-Design, Requirements-Test, Test-Code, Design-Code, and Design-Test. Among these artifact pairs, Issue-Commit, Issue-Issue, and Requirements-Code appeared in 71, 24, and 17 projects, respectively, making them the most frequently occurring types across our collected project corpus. In terms of research focus, Requirements-Code, Issue-Commit, and Requirements-Requirements were the most extensively studied artifact pairs, serving as research subjects in 24, 14, and 7 studies, respectively. We further analyzed the dataset from three dimensions — number of source artifacts, number of target artifacts, and size of the ground truth — and found that Issue-Commit, Issue-Issue, and Requirements-Code also ranked highest in terms of data scale. Notably, \textit{automation club} is the smallest project in the dataset, while \textit{ambari} is the largest project.

These projects serve as the dataset for the subsequent research in this paper. We have curated and organized these projects and have made them publicly available at \cite{source_code} for replication purpose.


\tcbset{
  colback=gray!10,    
  colframe=gray!80,   
  boxrule=0.5pt,      
  arc=2mm,            
  left=6pt, right=6pt, top=6pt, bottom=6pt, 
}

\begin{tcolorbox}[title=Observation 1]
In software TLR tasks over the past five years, Requirements-Code and Issue-Commit TLR have been the most widely studied tasks and are associated with the largest number of open-source projects.
\end{tcolorbox}
\FloatBarrier

\subsubsection{Answer RQ2: How effective is traceability link generation based on textual similarity across different types of project artifacts? }

\begin{table}[h]
\centering
\caption{Evaluation results of the ``Different Ratio'' metric and the effectiveness of textual similarity-based methods in the TLR task (``DR Rank'' indicates the ranking of the ``Different Ratio'' metric across different artifact types, and ``AF Rank'' indicates the ranking of the average F1-score across different artifact types).}
\label{DR_table}
\renewcommand{\arraystretch}{1.2} 
\begin{tabularx}{\textwidth}{>{\raggedright\arraybackslash}p{0.30\textwidth}
                             >{\raggedright\arraybackslash}p{0.11\textwidth}
                             >{\raggedright\arraybackslash}p{0.10\textwidth}
                             >{\raggedright\arraybackslash}p{0.11\textwidth}
                             >{\raggedright\arraybackslash}p{0.10\textwidth}
                             >{\raggedright\arraybackslash}p{0.07\textwidth}}
\toprule
\textbf{Artifacts types} & \textbf{Different ratio} & \textbf{DR rank} & \textbf{Avg F1} & \textbf{AF rank} & \textbf{p-value} \\
\midrule
Requirements-Code & +93.86\% & 8 & 0.3824 & 6 & \multirow{9}{*}{0.01} \\
Issue-Issue & +309.46\% & 3 & 0.4601 & 3 & \\
Design-Code & +258.38\% & 4 & 0.3952 & 4 & \\
Requirements-Requirements & +126.44\% & 7 & 0.3534 & 8 \\
Requirements-Design & +170.86\% & 5 & 0.3907 & 5 & \\
Test-Code & +885.19\% & 1 & 0.5602 & 1 & \\
Requirements-Test & +86.00\% & 9 & 0.3336 & 9 & \\
Issue-Commit & +465.19\% & 2 & 0.5602 & 2 & \\
Design-Test & +141.36\% & 6 & 0.3682 & 7 \\

\bottomrule
\end{tabularx}
\end{table}

To investigate the impact of artifact types on the effectiveness of text similarity-based methods in the TLR task, we analyzed the “Different Ratio” metric across various artifact type combinations. As shown in Table \ref{table app B} of Appendix B, the Different Ratio varies significantly among different artifact pairs, following the descending order: \textbf{Test-Code $>$ Issue-Commit $>$ Issue-Issue $>$ Requirements-Design $>$ Design-Test $>$ Requirements-Requirements $>$ Requirements-Code $>$ Requirements-Test}.

To further verify the relationship between the Different Ratio and the effectiveness of text similarity-based methods in TLR, we conducted a statistical significance test. As reported in Table \ref{DR_table}, the \textit{p-value} was $0.01 < 0.05$, allowing us to reject the null hypothesis $H_2$. This confirms a statistically significant correlation between the Different Ratio and the effectiveness of text similarity-based approaches in TLR. Moreover, since computing the Different Ratio requires significantly fewer computational resources compared to executing full-scale similarity-based link recovery, it can serve as a lightweight pre-evaluation indicator for determining whether a given TLR context is suitable for such methods.

\begin{tcolorbox}[title=Observation 2]
Traceability links between NL-NL and PL-PL artifacts (e.g., Test-Code, which involves code-related artifacts, or Issue-Commit, which involves artifacts described in NL) typically exhibit a higher Difference Ratio. In contrast, traceability links between NL and PL artifacts (e.g., Requirements-Code or Requirements-Test) generally have a lower Difference Ratio.
\end{tcolorbox}

The reason for this phenomenon is that NL-NL and PL-PL artifact pairs tend to be more structurally and semantically similar, making them more likely to share textual features such as identifiers and domain-specific terms. As a result, they are easier to distinguish using text similarity-based methods. In contrast, a semantic gap exists between NL and PL artifacts, which significantly reduces the discriminative power of text-based similarity methods.

In addition, we examined the consistency of this correlation across individual projects. Among the 99 projects analyzed, 97 had p-values below 0.05, indicating that in the vast majority of cases, text similarity-based methods are effective. However, two projects — \textit{smos} and \textit{quakus} — showed poor performance in terms of the Different Ratio. We hypothesize that in the \textit{smos} project, the high density of true links and tight coupling among code artifacts led to a negative Different Ratio and a \textit{p-value} of 0.9009. Similarly, in the \textit{quakus} project, the commit messages were overly brief and included a large amount of redundant information, resulting in minimal textual distinction between true and false links; hence the Different Ratio was also negative, with a \textit{p-value} of 1.0000.

These exceptional cases indicate that the textual quality and style of artifacts can significantly affect the effectiveness of text similarity-based approaches. In well-structured projects with clear and informative textual content, such methods are more likely to succeed. On the other hand, in projects with sparse descriptions or excessive noise, additional strategies - such as semantic modeling, structural analysis, or learning-based techniques - may be necessary to achieve satisfactory results.

\begin{tcolorbox}[title=Observation 3]
The Difference Ratio is a valuable metric for assessing whether text similarity-based methods are suitable for a given TLR task. Such methods perform well in TLR tasks involving NL-NL and PL-PL artifacts but experience a significant drop in effectiveness when applied to NL-PL artifact pairs.
\end{tcolorbox}

\subsubsection{Answer RQ3: What strategies can support text similarity-based methods in generating more accurate trace links for different types of artifacts?}
We conducted a statistical analysis of the strategies identified in the initial literature set collected in Section \ref{sec3.5}. The results of this analysis are presented in Table \ref{strategies}. The strategies and corresponding examples are explained as follows.
\begin{table}[h]
\centering
\caption{Statistics of traceability link recovery strategies in the initial literature set}
\label{strategies}
\renewcommand{\arraystretch}{1.2} 
\begin{tabularx}{\textwidth}{>{\raggedright\arraybackslash}p{0.05\textwidth}
>
{\raggedright\arraybackslash}p{0.35\textwidth}
                             >{\raggedright\arraybackslash}p{0.30\textwidth}
                             >{\raggedright\arraybackslash}p{0.15\textwidth}
                        }
\toprule
\textbf{No.}&\textbf{Strategy name} & \textbf{Artifacts type} & \textbf{Reference}  \\
\midrule
1&Leveraging intermediate artifacts & Design-Code; Requirements-Code; Requirements-Requirements& \cite{keim2024recovering, rodriguez2021leveraging, gao2024triad}\\
\midrule
2&Traceability link propagation for homogeneous artifacts & Requirements-Code; Requirements-Design; Requirements-Requirements; Issue-Commit; Issue-Issue & \cite{wang2021analyzing, zhu2022enhancing, tian2023cross} \\
\midrule
3&Date overlap score & Issue-Commit & \cite{yasa2025evaluating, zhu2024deep}\\
\midrule
4&Fine-grained & Requirements-Code; Issue-Commit; Design-Code & \cite{zhang2021recovering, peng2023enhancing, li2020combining, khlif2022complete, yoo2024building, hammoudi2021tracerefiner, hey2021improving, bai2024improving, shen2021supporting}\\
\midrule
5&Naming convention & Test-Code & \cite{sun2024method, white2022tctracer}\\
\midrule
6&Last call before assert & Requirements-Test& \cite{white2022tctracer}\\
\midrule
7&Code dependency & Requirements-Code & \cite{gao2022propagating, chen2021self, zou2024xwcode, zou2024enhancing}\\
\midrule
8&User feedback & Requirements-Code; Requirements-Requirements; Requirements-Test; Test-Code; Design-Test & \cite{gao2022propagating, du2020automatic}\\
\bottomrule
\end{tabularx}
\end{table}

\begin{enumerate}
    \item \textbf{Leveraging intermediate artifacts:} Due to the differences in expression between NL texts and PL texts, a significant semantic gap often exists in tasks such as recovering traceability links between requirements and code. To address this challenge, researchers have proposed an intermediate artifact strategy. This approach leverages an artifact that is semantically close to both the source and target artifacts to facilitate traceability link recovery.

   For example, in the task of recovering requirements-code traceability links, UML diagrams can be used as intermediate artifacts. UML diagrams can partially reflect the business logic and functional architecture described in the requirements, while also closely relating to the module structure and interaction patterns in the code implementation.

   \textbf{A concrete example is as follows}: In the requirements-to-code traceability link recovery task of the \textit{Dronology} project, the source artifact \textsc{RE-691.txt} (``RE'' refers to ``Requirements'') describes the system functions related to UAV operations, while \textsc{AFInfoBox.java} outlines the information generated from UAV activities. However, \textsc{RE-691.txt} tends to focus on the abstract concept of “UAV operation”, whereas \textsc{AFInfoBox.java} emphasizes implementation details, resulting in few shared terms between the two. In contrast, both \textsc{DD-692.txt} (``DD'' refers to ``Design Definition'') and \textsc{RE-691.txt}, as well as \textsc{DD-692.txt} and \textsc{AFInfoBox.java}, contain a greater number of shared terms. Therefore, this strategy uses \textsc{DD-692.txt} as an intermediate artifact, establishing the traceability link from the source artifact \textsc{RE-691.txt} to the target artifact \textsc{AFInfoBox.java} through the path \textsc{RE-691.txt} → \textsc{DD-692.txt} → \textsc{AFInfoBox.java}. 
    \item \textbf{Traceability link propagation for
homogeneous artifacts}: In the field of software traceability, similar artifacts often exhibit analogous link relationships. Taking the task of recovering requirements-code traceability links as an example, when developers implement code for a specific requirement, the corresponding code artifacts often show strong textual similarity. Based on this observation, software traceability researchers suggest that if two target artifacts exhibit a high degree of textual similarity, and one of them has already been explicitly linked to a given source artifact, it can be inferred that the other target artifact is also likely to be linked to the same source artifact.

\textbf{A concrete example is as follows:} In the requirements-to-code traceability link recovery task of the \textit{iTrust} project, \textsc{UC1.txt} represents the ``Create and Deactivate Patients'' use case, and both \textsc{AddPatientAction.java} and \textsc{AddPatientValidator.java} have traceability links to \textsc{UC1.txt}. The business process code in \textsc{AddPatientAction.java} is the implementation of the main flow described in \textsc{UC1.txt}, and since there are many shared terms, it is easier to identify the traceability link using text similarity-based methods. In contrast, the business logic in \textsc{AddPatientValidator.java} corresponds to the preconditions described in \textsc{UC1.txt}, but since the description of preconditions in \textsc{UC1.txt} is relatively limited, it is difficult to capture the link using text similarity-based methods. However, since the implementation of \textsc{AddPatientAction.java} is based on \textsc{AddPatientValidator.java}, and their textual content is relatively similar, it can be inferred that \textsc{AddPatientAction.java} and \textsc{AddPatientValidator.java} are highly likely to be related. Therefore, through the information transfer from \textsc{UC1.txt} → \textsc{AddPatientAction.java} → \textsc{AddPatientValidator.java}, it can be deduced that a traceability link is also likely to exist between \textsc{UC1.txt} and \textsc{AddPatientValidator.java}.

\item \textbf{Date overlap score:} In the task of recovering traceability links between issues and commits, if the submission time of a commit falls within the active time interval of an issue, this indicates a potential valid traceability link. On the contrary, if the submission time does not fall within the active time interval, the likelihood of a true traceability link between the commit and the issue is very low.

\textbf{A concrete example is as follows:} In the Issue-Commit traceability recovery task of the \textit{Tika} project, Issue \textsc{TIKA-1586} was created on \text2015-03-28 at 14:49:09 and closed on 2015-03-28 at 16:34:37. The commit with SHA ``a606c944c0ca4379cb6ab658b14b0f9a43bfbe6e'' was made on 2015-03-28 at 16:32:47, which falls within the active time interval of the issue. Moreover, the textual similarity between this commit and the issue is high, indicating the existence of a traceability link. In contrast, the commit with SHA ``4ae33b70378e0ee66b7d9c95d6fd7d51b10cc658'' was made on 2015-03-31 at 01:54:40, which is outside the active time interval of the issue; therefore, no traceability link exists between this commit and the issue.

\item \textbf{Fine-grained:} This strategy improves the accuracy of traceability link recovery by introducing more fine-grained levels of information. Taking the requirements–code traceability link recovery task as an example, conventional approaches often use the entire file as the comparison unit, which can easily overlook cases where only certain local structures within a code artifact exhibit strong semantic relevance to the requirement. In contrast, the fine-grained strategy decomposes the code into smaller components such as class names, class comments, method names, method comments, method parameters, and return types. It then calculates the textual similarity between each of these fragments and the requirement artifact. This allows the strategy to accurately identify potential traceability links even when the overall code artifact shares few terms with the requirement, by leveraging the strong alignment between local code structures and the content of the requirement.

\textbf{A concrete example is as follows:} In the requirements-code traceability link recovery task of the \textit{SMOS} project, a traceability link exists between \textsc{SMOS1.txt} and \textsc{User.java}. The file \textsc{SMOS1.txt} describes a login-related requirement, while \textsc{User.java}, serving as an entity class, contains a variety of attributes and methods. The overall textual content of \textsc{User.java} shows limited relevance to the functionality described in \textsc{SMOS1.txt}. However, by comparing only the method names with the functional requirement, a significant amount of implementation noise unrelated to the login functionality can be eliminated, which increases the likelihood of identifying a traceability link between \textsc{SMOS1.txt} and \textsc{User.java} based on textual similarity.

\item \textbf{Naming convention:} In the task of traceability link recovery between test code and production code, this strategy performs structural analysis on the names of test methods by splitting them into a prefix and a “root word”. It first determines whether the prefix belongs to a set of common test-related identifiers, such as “test”, “tests”, “testcase”, or “testcases”. The extracted root word is then compared with method names in the production code. If the prefix matches the predefined patterns and the root word corresponds to a method name in the production code, a potential traceability link is considered to exist between the test and the associated code component.

\textbf{A concrete example is as follows:} In the Test-Code traceability link recovery task for the \textit{Gson} project, the test method \textsc{testIsSynthetic()} and the production method \textsc{isSynthetic()} can be linked based on naming convention. This is because the test method has the prefix ``test'' and its root word ``isSynthetic'' matches the name of the production method, thereby establishing a potential traceability link through the naming convention strategy.

\item \textbf{Last call before assert:} This strategy is designed to establish traceability links between test code and production code by assuming that the function returned immediately before an assert statement is the target function being tested.

\textbf{A concrete examp;e is as folows:} Taking the Test-Code traceability link recovery task in the \textit{gson} project as an example, in the function \textsc{testEscapedBackslashInStringDeserialization()}, there is a statement ``\textsc{String actual = gson.fromJson(``'a\textbackslash\textbackslash b''', String.class);} \textsc{assertThat(actual).isEqualTo(``a\textbackslash\textbackslash b'');}'', where the function returned before the assert statement is \textsc{fromJson}. Therefore, it can be inferred that there exists a traceability link between \textsc{testEscapedBackslashInStringDeserialization()} and \textsc{fromJson(String, Class)}.

\item \textbf{Code dependency:} This strategy assists in recovering traceability links by leveraging structural relationships among code artifacts. In the context of requirements-to-code traceability recovery tasks, typical code dependencies include class inheritance (\textsc{extends}), interface implementation (\textsc{implements}), and package import (\textsc{import}). If a valid traceability link has already been established between a requirement artifact and a code artifact, other code artifacts that are directly connected to the linked artifact through such structural dependencies may also contribute to fulfilling the same requirement and therefore have a potential traceability relationship with the requirement artifact.

\textbf{A concrete example is as follows:} In the traceability link recovery task of the \textit{eTour} project, there exist traceability links between \textsc{UC2.txt} and both \textsc{CulturalHeritageAgencyManager.java} and \textsc{CulturalHeritageChecker.java}. The primary functionality of \textsc{UC2.txt} is to allow an agency operator to insert a new cultural heritage object into the system. Correspondingly, \textsc{CulturalHeritageAgencyManager.java} is mainly responsible for handling the interaction between cultural heritage objects and the database. This class also contains the \textsc{insertCulturalHeritage} method, which directly implements the main function described in \textsc{UC2.txt}. As a result, there is a high degree of textual similarity between \textsc{UC2.txt} and \textsc{CulturalHeritageAgencyManager.java}, facilitating the establishment of a traceability link between them. On the other hand, \textsc{CulturalHeritageChecker.java} primarily focuses on verifying the completeness of cultural heritage object data. Although this functionality is necessary for fulfilling the requirements specified in \textsc{UC2.txt}, the textual similarity between \textsc{CulturalHeritageChecker.java} and \textsc{UC2.txt} is very low. However, since \textsc{CulturalHeritageAgencyManager.java} has an ``import'' dependency on \textsc{CulturalHeritageChecker.java}, it is possible to infer a potential traceability link between \textsc{UC2.txt} and \textsc{CulturalHeritageChecker.java} through the existing links: \textsc{UC2.txt} → \textsc{CulturalHeritageAgencyManager.java} and \textsc{CulturalHeritageAgencyManager.java} → \textsc{CulturalHeritageChecker.java}.

\item \textbf{User feedback:} This strategy introduces a small set of high-confidence, manually annotated links provided by users to guide and enhance the automated traceability link recovery process. In real-world software development, fully relying on automated methods often fails to achieve optimal accuracy, whereas developers or domain experts are typically able to identify a subset of high-quality requirements-code links. The strategy leverages these manually confirmed samples as supervision signals to help the model learn the underlying semantic features and structural patterns, thereby generalizing to large-scale, unlabeled link prediction tasks.

\textbf{A concrete example is as follows:} Taking the traceability link recovery task in the \textit{iTrust} project as an example, the main function of \textsc{UC1.txt} is to create and deactivate patients, while \textsc{AddPatientValidator.java} is primarily responsible for validating the format of patient information. In contrast, \textsc{UC2.txt} focuses on creating, disabling, and editing personnel, and \textsc{AddPersonnelValidator.java} is used to validate the format of hospital personnel information. If the user provides feedback indicating a confirmed link between \textsc{UC1.txt} and \textsc{AddPatientValidator.java}, this high-confidence link can help the model identify similar links such as the one between \textsc{UC2.txt} and \textsc{AddPersonnelValidator.java}.

\begin{table}[h]
\centering
\caption{Details of strategy usage for different types of artifacts}
\label{strategy_usage}
\renewcommand{\arraystretch}{1.2} 
\begin{tabularx}{\textwidth}{>{\raggedright\arraybackslash}p{0.3\textwidth}
                             >{\raggedright\arraybackslash}p{0.20\textwidth}
                             >{\raggedright\arraybackslash}p{0.20\textwidth}
                             >{\raggedright\arraybackslash}p{0.15\textwidth}
                        }
\toprule
\textbf{Artifacts Type} & \textbf{Use Strategy} & \textbf{Not use strategy} & \textbf{Ratio} \\
\midrule
Requirements-Code & 14 & 24 & 58.33\%\\
Requirements-Requirements & 4 & 7 & 57.14\%\\
Requirements-Design & 3 & 8 & 37.50\% \\
Requirements-Test & 3 & 5 & 60.00\%\\
Issue-Commit & 5 & 17 & 29.41\% \\
Issue-Issue & 1 & 2 & 50.00\% \\
Design-Code & 2 & 2 & 100.00\%\\
Design-Test & 1 & 2 & 50.00\% \\
Test-Code & 2 & 2 & 100.00\%\\
\bottomrule
\end{tabularx}
\end{table}
\end{enumerate}

In addition, based on the data from Table \ref{strategies} and Table \ref{table app A} of Appendix A, we further analyzed the number of studies in the initial literature set involving different types of artifact pairs, as well as the number of studies that adopted enhancement strategies. The results are shown in Table \ref{strategy_usage}. It is observed that, on average, 54.81\% of the studies focusing on homogeneous artifact pairs (e.g., natural text–natural text or code text–code text) adopted the enhancement strategies described in Table \ref{strategies}. In contrast, for heterogeneous artifact pairs (e.g., natural text–code text), this proportion increases to 67.08\%. Combined with the findings in Table \ref{DR_table} and Table \ref{table app A} of Appendix A, we derive the following observation:

\begin{tcolorbox}[title=Observation 4]
For traceability link recovery tasks involving NL-NL or PL-PL artifact pairs, text similarity-based methods perform poorly, and therefore researchers are more likely to adopt enhancement strategies. In contrast, for NL-PL artifact pairs, text similarity-based methods perform well, making such strategies less necessary.
\end{tcolorbox}

\section{Case Study} \label{sec case study}
According to Observation 1, we select the requirements-to-code TLR task as a representative task for NL-PL TLR and conducts a case study, attempting to integrate various strategies mentioned in RQ3 using heterogeneous graph neural networks and prompt-based large language models. The goal of the case study is to assess whether these methods can effectively combine the information from different strategies and improve the performance of text similarity-based approaches.

\subsection{Baselines} \label{baselines}
In the task of requirements-code TLR, IR-based methods and learning-based methods are the two most widely adopted approaches \cite{aung2020literature}. Given that HGT is inherently a learning-based method, and that IR techniques represent the most commonly used text similarity approaches in this domain, we selected three representative IR and learning-based methods as experimental baselines to comprehensively compare their performance. The baseline methods are described as follows:

\begin{itemize}
    \item \textbf{TAROT} \cite{gao2022using}: A high-performance IR method that enhances the corpus using co-occurrence bigrams. It generally outperforms other IR methods on all projects.
    \item \textbf{GA-XWCoDe} \cite{zou2024enhancing}: This method combines the XGBoost classifier \cite{chen2016xgboost} with a genetic algorithm \cite{holland1992genetic} for automated parameter tuning. It achieves the best performance on the iTrust and SMOS projects.
    \item \textbf{HGNNLink} \cite{wang2025HGNNLink}: Our previously proposed method, currently the state-of-the-art in this domain. It is built upon the HGT framework and integrates a code dependency strategy, achieving top performance on all projects except iTrust and smos.

\end{itemize}

\subsection{Research Questions}
In order to explore whether certain methods can effectively integrate the various strategies described in RQ3 and outperform text similarity-based approaches, while also verifying the performance advantages of the proposed methods over existing research, this section will present two research questions. The rationale for each research question and the experimental design will be detailed in the following section.

\subsubsection{RQ4: Is there any method that can integrate multiple strategies to improve the performance of text similarity-based methods? } \label{RQ4}

Based on Observation 4, researchers tend to propose various strategies to address the limitations of text similarity-based methods for NL-PL artifacts. However, although different types of strategies have been employed to enhance the performance of text similarity-based methods from various perspectives, few studies have explored whether these strategies can be integrated to further improve model performance. Therefore, we pose \textit{RQ4: Is there a method that can integrate multiple strategies to improve the performance of text similarity-based methods?} 

To address this RQ, we select the requirements-to-code traceability link recovery (TLR) task — a representative NL-PL TLR task — as the subject of our study, aiming to explore whether integrating multiple strategies can significantly improve model performance when text similarity-based methods perform poorly. Based on the research of Zou et al. \cite{zou2024hantracer} and Bai et al. \cite{bai2024improving}, we believe that heterogeneous graph neural networks have the potential to capture multi-dimensional strategies in software TLR tasks. The information contained in various strategies can be constructed through edge relationships and transmitted within the graph structure. Additionally, according to the work of Deng et al. \cite{deng2024promptlink}, Wang et al. \cite{wang2025mplinker} and Fuchss et al. \cite{fuchss2025lissa}, we argue that prompt-based large language models also have the potential to integrate multi-dimensional strategy information. These models can transform strategy information into fixed-format text, supplementing the prompt description and helping the large language model obtain more accurate links. Moreover, HGT and prompt-based LLM represent the most commonly used supervised and unsupervised learning approaches, respectively, in current TLR tasks, and they broadly cover the mainstream methodological paradigms adopted in recent research. The design of heterogeneous graph neural networks and prompt-based large language models will be described in detail in Section \ref{methodology}. 

For the selection of the heterogeneous graph neural network, we referred to our previous research and chose the Heterogeneous Graph Transformer (HGT) \cite{hu2020heterogeneous} as the base model. We used RoBERTa \cite{liu2019roberta} to vectorize the requirement artifacts and GraphCodeBERT \cite{guo2020graphcodebert} to vectorize the code artifacts. The resulting vectors were used as node features for requirements and code artifacts, respectively. For edge feature construction, we built supervised edges based on the ground truth links, while message-passing edges were constructed based on the strategy relationships proposed in RQ3. Specifically, for the requirements-to-code TLR) task, RQ3 proposed five strategies. In this study, we selected three of them for in-depth analysis: “Fine-Grained”, “Code-Dependency”, and “User Feedback”. The “Leveraging Intermediate Artifacts” strategy was not adopted, as TLR datasets for requirements to code typically do not include third-party artifacts, making it difficult to implement in this context. The “Traceability Link Propagation for Homogeneous Artifacts” strategy is generally applied during the link refinement stage (i.e., after candidate links have been generated by the model). However, the method proposed in this study integrates strategies during the link generation phase, rendering this strategy inapplicable to our case study.

For the LLM model, based on the preliminary experimental results shown in Table \ref{LLM_performance}, we selected the Gemini 2.5 Pro model \cite{team2023gemini}. The strategies used are consistent with those applied in the heterogeneous graph neural network approach. The specific application of these strategies and the construction of prompt templates will be detailed in Section \ref{gemini-all}.

\begin{table}[htbp]
\centering
\caption{Comparison of Large Language Model Performance on the Requirements-to-Code Traceability Link Recovery Task (Three gray background colors, from light to dark, are used to highlight the top three results (Third, Second, First))}
\label{LLM_performance}
{\footnotesize
\begin{tabularx}{\textwidth}{%
  p{1.5cm}        
  p{2cm} 
  p{1cm}        
  p{1cm}        
  p{1.2cm}        
  p{2cm} 
  p{1cm}        
  p{1cm}        
  p{1.2cm}        
}
\toprule
Project & Model & Precision& Recall & F1-Score & Model & Precision & Recall & F1-Score \\
\midrule
\multirow{4}{*}{Albergate} & Gemini 2.5 Pro& \cellcolor{gray!40}0.8000 & 0.6667 & \cellcolor{gray!20}0.7273 & Llama 3.3 8B& 0.7500 & \cellcolor{gray!70}1.0000 & \cellcolor{gray!70}0.8571 \\
& Devstral small & 0.7143 & \cellcolor{gray!40}0.8333 & \cellcolor{gray!40}0.7692 & Deepseek-R1 & 0.6000 & 0.5000 & 0.5455\\
&Deepcoder 14B & 0.6250 & \cellcolor{gray!40}0.8333 & 0.7143 & Deepseek-V3 & \cellcolor{gray!40}0.8000 & 0.6667 & \cellcolor{gray!20}0.7273 \\
& Codex Mini & 0.7500 & 0.5000 & 0.6000 & Coder Large & \cellcolor{gray!70}1.0000 & 0.5000 & 0.6667\\
\midrule
\multirow{4}{*}{Derby} & Gemini 2.5 Pro& 0.9362 & \cellcolor{gray!70}0.6217 & \cellcolor{gray!70}0.7472 & Llama 3.3 8B& 0.6628 & \cellcolor{gray!20}0.5044 & 0.5729 \\
& Devstral small & 0.6685 & \cellcolor{gray!40}0.5442 & \cellcolor{gray!40}0.6000 & Deepseek-R1 & 0.9333 & 0.4336 & 0.5921\\
&Deepcoder 14B & 0.8387 & 0.4602 & \cellcolor{gray!20}0.5943 & Deepseek-V3 & \cellcolor{gray!20}0.9444 & 0.1504 & 0.2595 \\
& Codex Mini &\cellcolor{gray!40} 0.9595 & 0.3142 & 0.4733 & Coder Large & \cellcolor{gray!70}1.0000 & 0.1549 & 0.2682\\
\midrule
\multirow{4}{*}{Dronology} & Gemini 2.5 Pro& 0.5745 & \cellcolor{gray!70}0.6000 & 0.5870 & Llama 3.3 8B& 0.7429 & \cellcolor{gray!20}0.5778 & \cellcolor{gray!40}0.6500 \\
& Devstral small & 0.7097 & 0.4889 & 0.5789 & Deepseek-R1 & 0.7692 & 0.4444 & 0.5634\\
&Deepcoder 14B & 0.7941 & \cellcolor{gray!70}0.6000 & \cellcolor{gray!70}0.6835 & Deepseek-V3 & \cellcolor{gray!20}0.8696 & 0.4444 & \cellcolor{gray!20}0.5882 \\
& Codex Mini & \cellcolor{gray!70}0.9500 & 0.4222 & 0.5846 & Coder Large & \cellcolor{gray!40}0.8824 & 0.3333 & 0.4839\\
\midrule
\multirow{4}{*}{Drools} & Gemini 2.5 Pro& 0.8333 & \cellcolor{gray!70}0.5275 & \cellcolor{gray!70}0.6460 & Llama 3.3 8B& 0.7119 & \cellcolor{gray!40}0.4884 & \cellcolor{gray!40}0.5793 \\
& Devstral small & 0.5962 & \cellcolor{gray!20}0.3605 & 0.4493 & Deepseek-R1 & 0.8438 & 0.3140 & 0.4576\\
&Deepcoder 14B & 0.8108 & 0.3488 & \cellcolor{gray!20}0.4878 & Deepseek-V3 & \cellcolor{gray!40}0.8889 & 0.0093 & 0.1684 \\
& Codex Mini & \cellcolor{gray!20}0.8696 & 0.2326 & 0.3670 & Coder Large & \cellcolor{gray!70}1.0000 & 0.1163 & 0.2083\\
\midrule
\multirow{4}{*}{eAnci} & Gemini 2.5 Pro& 0.5490 & \cellcolor{gray!70}0.4912 & \cellcolor{gray!70}0.5185 & Llama 3.3 8B& 0.5417 & \cellcolor{gray!20}0.2281 & \cellcolor{gray!20}0.3210 \\
& Devstral small & 0.6875 & 0.1930 & 0.3014 & Deepseek-R1 & \cellcolor{gray!20}0.8750 & \cellcolor{gray!40}0.2456 & \cellcolor{gray!40}0.3836\\
&Deepcoder 14B & 0.6111 & 0.1930 & 0.2933 & Deepseek-V3 & \cellcolor{gray!70}1.0000 & 0.0526 & 0.1000 \\
& Codex Mini & \cellcolor{gray!70}1.0000 & 0.0702 & 0.1311 & Coder Large & 0.7500 & 0.0526 & 0.0984\\
\midrule
\multirow{4}{*}{Groovy} & Gemini 2.5 Pro& 0.8785 & \cellcolor{gray!70}0.6667 & \cellcolor{gray!70}0.7581 & Llama 3.3 8B& 0.7143 & 0.2778 & 0.4000 \\
& Devstral small & 0.5000 & \cellcolor{gray!40}0.4444 & 0.4706 & Deepseek-R1 & 0.8750 & 0.3889 & \cellcolor{gray!20}0.5385\\
&Deepcoder 14B & 0.7778 & 0.3889 & 0.5185 & Deepseek-V3 & \cellcolor{gray!70}1.0000 & 0.2778 & 0.4348 \\
& Codex Mini & \cellcolor{gray!70}1.0000 & \cellcolor{gray!40}0.4444 & \cellcolor{gray!40}0.6154 & Coder Large & \cellcolor{gray!70}1.0000 & 0.2222 & 0.3636\\
\midrule
\multirow{4}{*}{Infinispan} & Gemini 2.5 Pro& 0.9111 & \cellcolor{gray!70}0.5857 & \cellcolor{gray!70}0.7130 & Llama 3.3 8B& 0.8611 & 0.4429 & 0.5849 \\
& Devstral small & 0.8659 & \cellcolor{gray!40}0.5071 & \cellcolor{gray!40}0.6396 & Deepseek-R1 & \cellcolor{gray!40}0.9273 & 0.3643 & 0.5231\\
&Deepcoder 14B & 0.8750 & \cellcolor{gray!20}0.4500 & \cellcolor{gray!20}0.5943 & Deepseek-V3 & \cellcolor{gray!70}0.9444 & 0.2429 & 0.3864 \\
& Codex Mini & 0.9143 & 0.2286 & 0.3657 & Coder Large & \cellcolor{gray!20}0.9259 & 0.1786 & 0.2994\\
\midrule
\multirow{4}{*}{iTrust} & Gemini 2.5 Pro& 0.8696 & \cellcolor{gray!40}0.7407 & \cellcolor{gray!70}0.8000 & Llama 3.3 8B& \cellcolor{gray!70}1.0000 & 0.5926 & 0.7442 \\
& Devstral small & 0.7333 & \cellcolor{gray!70}0.8148 & 0.7719 & Deepseek-R1 & \cellcolor{gray!70}1.0000 & 0.6296 & \cellcolor{gray!20}0.7727\\
&Deepcoder 14B & 0.7917 & 0.7037 & 0.7451 & Deepseek-V3 & \cellcolor{gray!70}1.0000 & 0.4444 & 0.6154 \\
& Codex Mini & 0.8696 & \cellcolor{gray!40}0.7407 & \cellcolor{gray!70}0.8000 & Coder Large & 0.9375 & 0.5556 & 0.6977\\
\midrule
\multirow{4}{*}{maven} & Gemini 2.5 Pro& 0.7000 & \cellcolor{gray!70}0.7778 & \cellcolor{gray!70}0.7368 & Llama 3.3 8B& 0.6667 &\cellcolor{gray!70}0.7778 & \cellcolor{gray!20}0.7179 \\
& Devstral small & 0.7000 & \cellcolor{gray!70}0.7778 & \cellcolor{gray!70}0.7368 & Deepseek-R1 & \cellcolor{gray!70}1.0000 & 0.5000 & 0.6667\\
&Deepcoder 14B & \cellcolor{gray!70}1.0000 & 0.5556 & 0.7143 & Deepseek-V3 & \cellcolor{gray!70}1.0000 & 0.3889 & 0.5600 \\
& Codex Mini & \cellcolor{gray!70}1.0000 & 0.3333 & 0.5000 & Coder Large & \cellcolor{gray!70}1.0000 & 0.3333 & 0.5000\\
\midrule
\multirow{4}{*}{Pig} & Gemini 2.5 Pro& 0.6727 &\cellcolor{gray!70} 0.6271 & \cellcolor{gray!40}0.6491 & Llama 3.3 8B& 0.8222 &\cellcolor{gray!70} 0.6271 & \cellcolor{gray!70}0.7115 \\
& Devstral small & 0.5106 & 0.4068 & 0.4528 & Deepseek-R1 & \cellcolor{gray!40}0.9286 & 0.2203 & 0.3562\\
&Deepcoder 14B & 0.7778 & \cellcolor{gray!20}0.4746 & \cellcolor{gray!20}0.5895 & Deepseek-V3 & 0.8571 & 0.3051 & 0.4500 \\
& Codex Mini & \cellcolor{gray!20}0.8800 & 0.3729 & 0.5238 & Coder Large & \cellcolor{gray!70}0.9333 & 0.2373 & 0.3784\\
\midrule
\multirow{4}{*}{Seam2} & Gemini 2.5 Pro& 0.6538 & \cellcolor{gray!40}0.5235 & \cellcolor{gray!40}0.5814 & Llama 3.3 8B& 0.6538 & \cellcolor{gray!20}0.4250 & \cellcolor{gray!20}0.5152 \\
& Devstral small & 0.6857 & \cellcolor{gray!70}0.6000 & \cellcolor{gray!70}0.6400 & Deepseek-R1 & 0.6154 & 0.4000 & 0.4848\\
&Deepcoder 14B & 0.6538 & \cellcolor{gray!20}0.4250 & \cellcolor{gray!20}0.5152 & Deepseek-V3 & \cellcolor{gray!40}0.7500 & 0.3000 & 0.4286 \\
& Codex Mini & \cellcolor{gray!20}0.7368 & 0.3500 & 0.4746 & Coder Large &\cellcolor{gray!70} 1.0000 & 0.3000 & 0.4615\\
\midrule
\multirow{4}{*}{smos} & Gemini 2.5 Pro& 0.7714 & \cellcolor{gray!70}0.2500 & \cellcolor{gray!70}0.3776 & Llama 3.3 8B& 0.7714 & \cellcolor{gray!70}0.2500 & \cellcolor{gray!70}0.3776 \\
& Devstral small & 0.4706 & \cellcolor{gray!20}0.2222 & 0.3019 & Deepseek-R1 & 0.7083 & 0.1574 & 0.2576\\
&Deepcoder 14B & 0.6857 & \cellcolor{gray!20}0.2222 & \cellcolor{gray!20}0.3357 & Deepseek-V3 & \cellcolor{gray!70}0.9412 & 0.1481 & 0.2560 \\
& Codex Mini & \cellcolor{gray!40}0.8182 & 0.1667 & 0.2769 & Coder Large & \cellcolor{gray!40}0.8182 & 0.0833 & 0.1513\\
\bottomrule
\end{tabularx}
}
\caption*{ \footnotesize
Note: P denotes Precision, R denotes Recall. Gemini 2.5 Pro refers to Google: Gemini 2.5 Pro Preview 06-05,
Llama 3.3 8B refers to Meta: Llama 3.3 8B Instruct,
Devstral Small refers to Mistral: Devstral Small,
Deepseek-R1 refers to Deepseek: R1 0528,
Deepcoder 14B refers to Agentica: Deepcoder 14B Preview,
Deepseek-V3 refers to Deepseek: V3 0324,
Codex Mini refers to OpenAI: Codex Mini,
Coder Large refers to Arcee: Coder Large.}
\end{table}

For dataset selection, we filtered the requirements-code TLR projects obtained in RQ1. Since it is not possible to construct abstract syntax trees (ASTs) for code files containing syntax errors—thereby hindering the extraction of code dependency information—we excluded projects with a high number of syntax errors. As a result, we retained 12 open-source projects: Albergate, Derby, Dronology, Drools, eAnci, Groovy, Infinispan, iTrust, Maven, Pig, Seam2, and SMOS, which were used as the experimental dataset for this RQ.

To address the RQ, we designed and conducted an ablation study involving three categories of models. The first category includes baseline models without any enhancement strategies (HGT-None and Gemini-None). The second category consists of models that incorporate a single enhancement strategy, including HGT-Code Dependency, HGT-User Feedback, HGT-Fine Grained, and their corresponding Gemini variants. The third category includes models that combine all enhancement strategies (HGT-All and Gemini-All). By comparing models that apply a single strategy with those that apply none, we aim to assess whether these strategies help mitigate the limitations of relying solely on textual similarity methods. Furthermore, by comparing models that incorporate all strategies with those using only a single strategy, we explore whether HGT and Gemini, when combined with prompt-based learning, can effectively integrate multi-dimensional features and enhance the utilization of each feature. The evaluation metrics used in this study are Precision, Recall, and F1-score, as described in Section \ref{sec_metric}.

In addition, to determine whether HGT-All and Gemini-All perform significantly better than their corresponding single-strategy variants and original models, we apply the Wilcoxon Signed-Rank Test to conduct pairwise comparisons based on F1-score results across 12 projects. This non-parametric statistical test is suitable for paired sample data with small sample sizes where the assumption of normality may not hold, and it is widely used to evaluate statistically significant differences between two related samples.

For the HGT-All model, we define the following null and alternative hypotheses:
\begin{itemize}
    \item $H_4$: HGT-All shows no significant performance difference compared to its variants.
    \item $H_5$: HGT-All significantly outperforms its variants.
\end{itemize}
For the Gemini-All model, the hypotheses are:
\begin{itemize}
    \item Gemini-All shows no significant performance difference compared to its variants
    \item Gemini-All significantly outperforms its variants.
\end{itemize}

First, for each project, we compute the F1-score difference between the model pairs:
\begin{equation}
    d_i=x_i-y_i
\end{equation}
where $x_i$ represents the F1-score of HGT-All or Gemini-All, and $y_i$ is the F1-score of the corresponding variant. We exclude all samples where $d_i=0$, take the absolute values of the remaining differences, and rank them in ascending order to assign ranks $R_i$. We then separate the ranks into positive and negative groups based on the sign of $d_i$ and compute the sums:
\begin{equation}
    W^+=\sum_{d_i>0}R_i, \,\,\,W^-=\sum_{d_i<0}R_i
\end{equation}

The final test statistic is calculated as:
\begin{equation}
    W=min(W^+,W^-)
\end{equation}

Based on the test statistic $W$ and sample size $n$, we use the Python scipy library to retrieve the corresponding \textit{p-value} from statistical tables. If the \textit{p-value} is less than the significance level of $0.05$, we reject the null hypothesis and conclude that HGT-All or Gemini-All performs significantly better than their single-strategy variants from a statistical perspective.

\subsubsection{RQ5: How do HGT-All and Gemini-All perform in the requirements-code TLR task?}

To further investigate the performance advantages of HGT-All and Gemini-All compared to other baseline methods, we propose \textit{RQ5: How do HGT-All and Gemini-All perform in the requirements-code TLR task?}

To address this RQ, we compare HGT-All and Gemini-All with the baseline methods introduced in Section \ref{baselines}, using Precision, Recall, and F1-score as evaluation metrics. In addition, to assess whether the performance differences are statistically significant, we apply the same Wilcoxon Signed-Rank Test used in RQ4, based on the F1-scores of the models across 12 projects.

Specifically, for the HGT-All model, we define the following null and alternative hypotheses:
\begin{itemize}
    \item $H_8$: There is no significant performance difference between HGT-All and the baseline methods.
    \item $H_9$: HGT-All significantly outperforms the baseline methods.
\end{itemize}

Similarly, for the Gemini-All model, the hypotheses are defined as:
\begin{itemize}
    \item $H_{10}$: There is no significant performance difference between Gemini-All and the baseline methods.
    \item $H_{11}$: Gemini-All significantly outperforms the baseline methods.
\end{itemize}

\subsection{Methodology} \label{methodology}
In this section, we provide a detailed description of HGT-All and Gemini-All (explained in Section \ref{rq4}).
\subsubsection{HGT-All}
HGT-All includes three components: node construction, edge construction, and a link prediction process based on the Heterogeneous Graph Transformer (HGT). The structure of the heterogeneous graph is defined as $\mathcal{G}=(\mathcal{V},\mathcal{E})$, where $\mathcal{V}$ denotes the set of nodes and $\mathcal{E}$ denotes the set of edges.

In the node construction phase, we uniformly model requirement artifacts and code artifacts as two types of nodes in the heterogeneous graph. Specifically, given a requirement document $r_i$, we use the pre-trained language model RoBERTa to encode it and obtain the initial representation of the requirement node:
\begin{equation}
    h_{r_i}^{(0)}=RoBERTa(r_i)\in\mathbb{R}^d
\end{equation}

where $d$ denotes the dimensionality of RoBERTa’s output, which is 768. Similarly, for each code file $c_j$, we use GraphCodeBERT for encoding:
\begin{equation}
    h_{c_j}^{(0)}=GraphCodeBERT(c_j)\in\mathbb{R}^d
\end{equation}

The vectors above constitute the initial input representations of the heterogeneous graph nodes and serve as the layer-0 input for the HGT model.
\begin{figure}
    \centering
    \includegraphics[width=\linewidth]{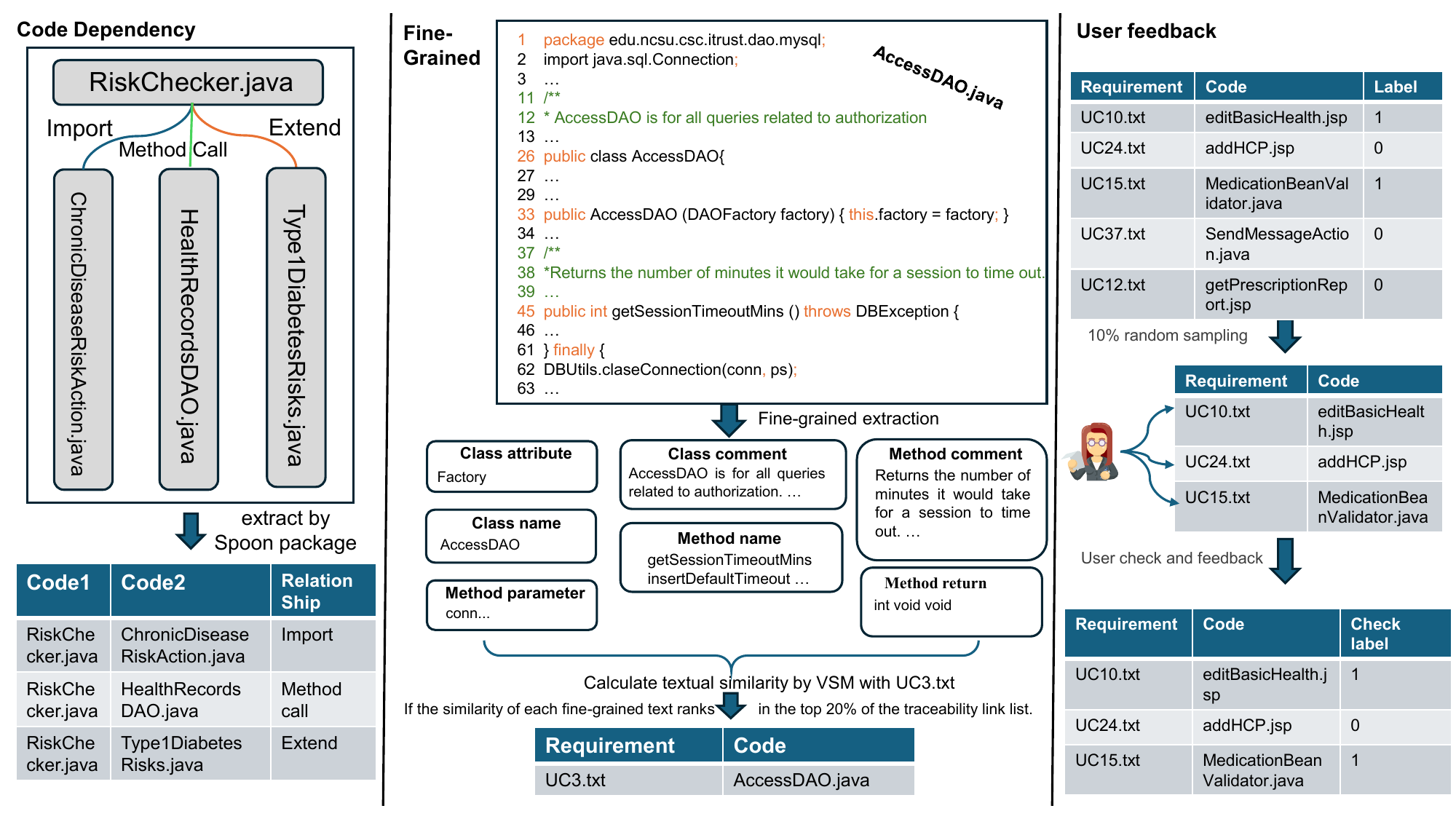}
    \caption{Illustration of Strategy Relationship Extraction for Code Dependency, Fine-Grained, and User Feedback (using the iTrust Project as an Example)}
    \label{strategy_relationship}
\end{figure}
In the edge construction phase, we introduce three types of strategy-driven edge relations, as shown in figure \ref{strategy_relationship}, and constructed as follows:
\begin{enumerate}
    \item \textbf{Code dependency edges:} We extract three types of structural dependencies from the source code of the project: \textsc{import}, \textsc{extend}, and \textsc{method call}. If there exists a dependency between code files $c_i$ and $c_j$ denoted as $\text{dep}_t(c_i,c_j)=1$, where $t\in {\text{import},\text{extend},\text{call}}$, we add an edge of type $t$ to the graph:
    \begin{equation}
        e_{ij}^t=(c_i,c_j)\in\mathcal{E}
    \end{equation}
    \item \textbf{User feedback edges:} To simulate real-world development scenarios where users provide partial feedback on trace links, we randomly sample $10\%$ of the links from the ground truth traceability set. For each sampled positive link $(r_i,c_j)$, we mark it as user feedback and add a feedback edge:
    \begin{equation}
        e_{ij}^{feedback}=(r_i,c_j)\in\mathcal{E}
    \end{equation}
    \item \textbf{Fine-grained semantic edges:} Each code file is decomposed into seven fine-grained components: \textit{class attribute}, \textit{class comment}, \textit{class name}, \textit{method comment}, \textit{method name}, \textit{method parameter}, \textit{method return}. Denote the text of each code component as $t_k(c_j)$. We use the Vector Space Model (VSM) to compute its similarity with requirement $r_i$:
    \begin{equation}
        sim_k(r_i,c_j)=cos(TFIDF(r_i),TFIDF(t_k(c_j)))
    \end{equation}
    If the following condition is satisfied:
    \begin{equation}
        \forall k\in\{1,...,7\},\,\,\,rank_{r_i}(t_k(c_j))\leq0.2*N_k
    \end{equation}
    meaning all components of $c_j$ rank within the top 20\% in similarity with $r_i$, then a fine-grained semantic edge is constructed:
    \begin{equation}
        e_{ij}^{fine\_grained}=(r_i,c_j)\in\mathcal{E}
    \end{equation}
\end{enumerate}

After the above construction, we obtain the heterogeneous graph $\mathcal{G}$, where node types include requirement nodes and code nodes, and edge types include five kinds: three types of code dependency relations, user feedback edges, and fine-grained semantic edges.

\begin{figure}
    \centering
    \includegraphics[width=\linewidth]{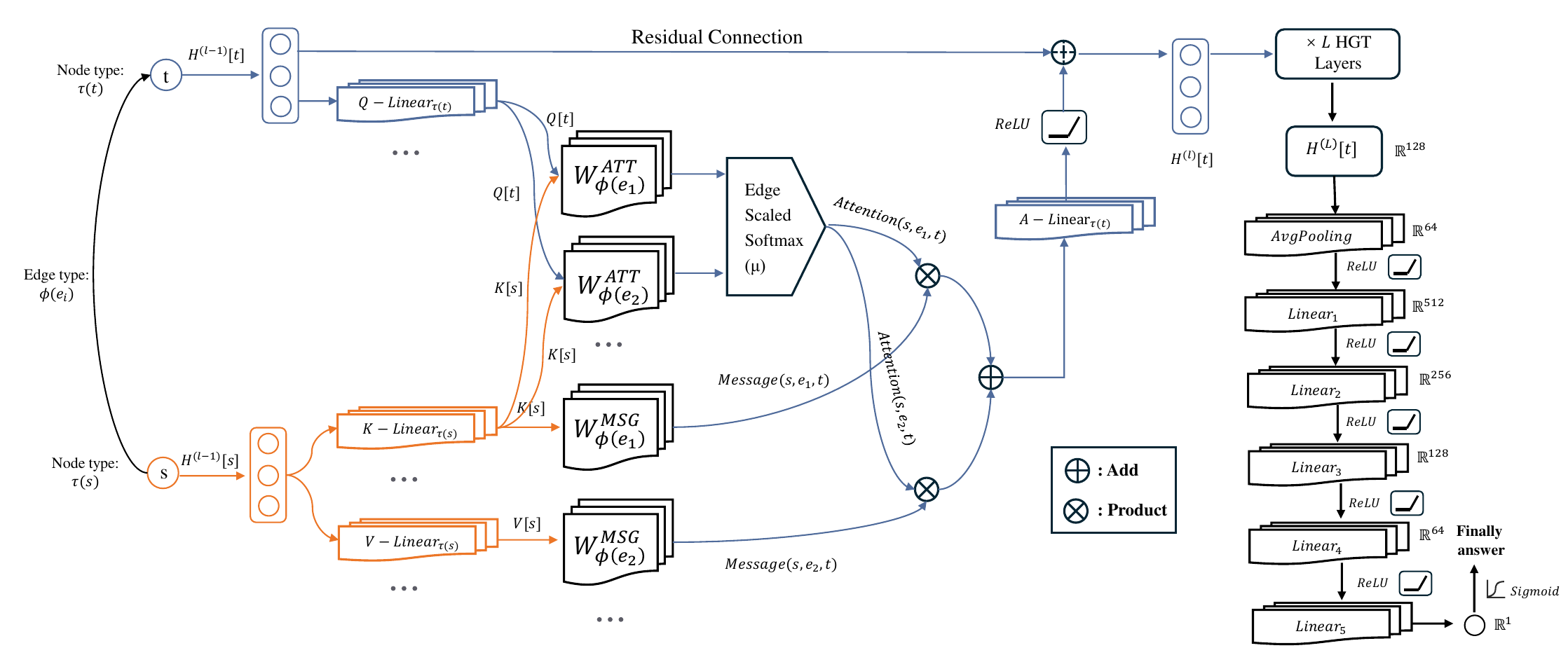}
    \caption{Flowchart of the requirements-to-code traceability link recovery using the HGT model}
    \label{flowchart_hgt}
\end{figure}

In the link prediction phase of the Heterogeneous Graph Transformer (HGT), each edge in the heterogeneous graph can be formally represented as a meta-relation triplet $<\tau(s),\Phi(e),\tau(t)>$, where $s$ and $t$ denote the source node (requirement artifact) and the target node (code artifact), respectively. $\tau(\cdot)$ is the node type mapping function, and $\Phi(e)$ indicates the edge type. The overall prediction process is illustrated in Figure \ref{flowchart_hgt}.

To improve training efficiency and reduce computational complexity, we introduce a type-aware neighbor sampling strategy at each HGT layer. For any target node $t$, we sample a fixed number of source nodes $s\in\mathcal{N}(t)$ from each of its different types of neighbors $\mathcal{N}(t)$ to construct a type-balanced local subgraph. This facilitates mini-batch training while preserving structural semantics and ensuring model scalability.

At the $l$-th HGT layer, the representation of each node $v$ is denoted as $h_v^{(l)}\in\mathbb{R}^d$, with input being the representation from the previous layer $h_v^{(l-1)}$. For any edge $e=(s,t)\in\mathcal{E}$, HGT first constructs a type-aware multi-head attention mechanism. In the $i$-th attention head, the target node $t$ is mapped to a query vector and the source node $s$ is mapped to a key vector:
\begin{equation}
    Q_i(t)=Linear_{i,\tau(t)}^Q(h_t^{(l-1)})
\end{equation}
\begin{equation}
    K_i(s)=Linear_{i,\tau(s)}^K(h_s^{(l-1)})
\end{equation}

where $Linear$ denotes a set of linear transformations. To model the influence of edge types on attention, an edge-type-specific transformation matrix $W_{\Phi(e)}^{ATT}\in\mathbb{R}^{d_h*d_h}$ is introduced, along with a scaling factor $\mu<\tau(s),\Phi(e),\tau(t)>\in \mathbb{R}$ to reflect the importance of the meta-relation triplet. The attention score for this head is computed as:
\begin{equation}
    ATT-Head_i(s,e,t)=\frac{Q_i(t)^\top W_{\Phi(e)}^{ATT}k_i(s)}{\sqrt{d_h}}\cdot\mu<\tau(s),\Phi(e),\tau(t)>
\end{equation}

Scores from all attention heads are concatenated and normalized over the neighbor set $\mathcal{N}(t)$ to yield the attention weight for edge $e=(s,t)$:
\begin{equation}
    \alpha_{s,e,t}=softmax_{s\in\mathcal{N}(t)}(||_{i=1}^h ATT-Head_i(s,e,t))
\end{equation}

During message passing, each source node $s$ is mapped to a value vector based on its type. This is then transformed using an edge-type-specific message transformation matrix $W_{\Phi(e)}^{MSG}\cdot Linear_{i,\tau(s)}^M(h_s^{(l-1)})$

All head messages are concatenated to form the multi-head message vector:
\begin{equation}
    m_{s,e,t}=||_{i=1}^h MSG-Head_i(s,e,t)
\end{equation}

Next, the target node $t$ aggregates the messages from all its neighbors weighted by the attention coefficients:
\begin{equation}
    \tilde{h_t}^{(l)}=\sum_{s\in\mathcal{N}(t)}\alpha_{s,e,t}\cdot m_{s,e,t}
\end{equation}

To enhance the expressiveness of node types, the aggregated result is passed through a node-type-specific linear transformation. It is then added via residual connection to the previous layer’s representation and activated by ReLU to get the updated node representation:
\begin{equation}
    h_t^{(l)}=\sigma(Linear_{\tau(t)}^A(\tilde{h_t}^{(l)}))+h_t^{(l-1)}
\end{equation}

After stacking two HGT layers, we obtain the final representation $h_v^{(L)}\in\mathbb{R}^{128}$ for each node, which incorporates semantic information from multi-type neighbors. We denote the representation of requirement nodes as $z_{r_i}$, and that of code nodes as $z_{c_j}$.

Next, we perform average pooling over each type of node and apply a linear transformation to compress the representations into 64-dimensional vectors:
\begin{equation}
    r_i=Linear(AvgPool(z_{r_i}))\in\mathbb{R}^{64}
\end{equation}
\begin{equation}
    c_j=Linear(AvgPool(z_{c_j}))\in\mathbb{R}^{64}
\end{equation}

These vectors are then input into a multi-layer feedforward neural network:
\begin{equation}
    h_{out}=f_4(f_3(f_2(f_1(r_i||c_j))))
\end{equation}

where each layer performs the following transformations: $f_1:\mathbb{R}^{128}\rightarrow\mathbb{R}^{512}$, $f_2:\mathbb{R}^{512}\rightarrow\mathbb{R}^{256}$, 
$f_3:\mathbb{R}^{256}\rightarrow\mathbb{R}^{128}$,
$f_4:\mathbb{R}^{128}\rightarrow\mathbb{R}^{64}$,$f_5:\mathbb{R}^{64}\rightarrow\mathbb{R}^1$. Each layer uses ReLU activation and Batch Normalization. Finally, the output representations of the requirement and code artifacts are used to compute an inner product, followed by a Sigmoid activation to obtain the link prediction probability:
\begin{equation}
    \hat{y_{i,j}}=\sigma(r_i^\top c_j)
\end{equation}

\subsubsection{Gemini-All} \label{gemini-all}
In the Gemini-All method, the strategy for obtaining prediction results is consistent with HGT-All. We first set the role of Gemini to system and assign it the following system-level instruction:
\begin{tcolorbox}[colback=blue!5, colframe=blue!75!black, title=System-level instruction]
You are a judge in the field of software traceability.
\end{tcolorbox}

Each requirements–code pair is constructed using a fixed prompt template in the following format:
\begin{tcolorbox}[colback=blue!5, colframe=blue!75!black, title=Prompt Template]
Determine if the following Requirements and Code are related. Answer only ``Yes'' or ``No''.

Requirements: \{requirements\_text\}

Code: \{code\_text\}

Additional Information: \{additional\_information\}
\end{tcolorbox}

The Additional Information section is automatically generated according to the following rules:
\begin{enumerate}
    \item If the current code artifact has structural relationships with other code artifacts in the project (e.g., extend, import, or method call), the following statement is added:
    \begin{tcolorbox}[colback=blue!5, colframe=blue!75!black, title=Additional\_Information 1]
{code1} and {code2} have a {relation} relationship.
\end{tcolorbox}
\item If user feedback exists for the current requirement–code pair, the following statement is added:
    \begin{tcolorbox}[colback=blue!5, colframe=blue!75!black, title=Additional\_Information 2]
User feedback indicates label is {0 or 1}.
\end{tcolorbox}
Otherwise, the following is added:
    \begin{tcolorbox}[colback=blue!5, colframe=blue!75!black, title=Additional\_Information 2']
No user feedback information.
\end{tcolorbox}
\item If there exists a fine-grained semantic similarity relationship between the requirement and the code, the following statement is added:
    \begin{tcolorbox}[colback=blue!5, colframe=blue!75!black, title=Additional\_Information 3]
Fine-grained relationship exists between {requirements} and {code}.
\end{tcolorbox}
Otherwise, the following is added:
    \begin{tcolorbox}[colback=blue!5, colframe=blue!75!black, title=Additional\_Information 3']
No fine-grained relationship between {requirements} and {code}.
\end{tcolorbox}
\end{enumerate}

Additionally, according to the official Gemini documentation\footnote{https://ai.google.dev/api/models}, this task falls under the data analysis category. Therefore, we set the temperature parameter to the recommended value of 1.0.

\subsection{Results}
\subsubsection{Answer RQ4: Is there any method that can integrate multiple strategies to improve the performance of text similarity-based methods?}
The experimental results for this RQ are shown in Table \ref{HGT_ablation} and Table \ref{gemini_ablation}. For HGT, we observe that incorporating different types of strategy-enhancing edges—namely, code dependency relations (average F1 = 0.6950), user feedback (average F1 = 0.6619), and fine-grained associations (average F1 = 0.6562)—consistently outperforms the baseline variant HGT-None, which does not employ any strategy-enhancing edges. This indicates that each of these strategies contributes positively to the performance of HGT by enriching the relational information in the graph, thereby compensating for the limitations of relying solely on textual similarity.

\begin{table}[htbp]
\centering
\caption{Ablation results of strategy edges in HGT on the Requirements-to-Code Traceability Link Recovery Task (Three gray background colors, from light to dark, are used to highlight the top three results (Third, Second, First))}
\label{HGT_ablation}
{\footnotesize
\begin{tabular}{llccclccc}
\toprule
Approach & Project & Precision & Recall & F1-Score & Project & Precision & Recall & F1-Score \\
\midrule
HGT-None & \multirow{5}{*}{Albergate} & 0.3973 & \cellcolor{gray!20}0.6000 & 0.4709 & \multirow{5}{*}{Derby} & \cellcolor{gray!20}0.7640 & \cellcolor{gray!20}0.7074 & \cellcolor{gray!20}0.7338 \\
HGT-Code dependency & & \cellcolor{gray!70}0.6667 & 0.4000 & \cellcolor{gray!20}0.5000 & & 0.7603 & 0.6645 & 0.7077 \\
HGT-User feedback & & \cellcolor{gray!40}0.5000 & \cellcolor{gray!20}0.6000 & \cellcolor{gray!40}0.5455 & & \cellcolor{gray!40}0.7664 & \cellcolor{gray!40}0.7100 & \cellcolor{gray!40}0.7371 \\
HGT-Fine grained & & 0.4189 & \cellcolor{gray!40}0.6200 & 0.4914 & & \cellcolor{gray!70}0.7744 & 0.6853 & 0.7255 \\
HGT-All & & \cellcolor{gray!20}0.4444 & \cellcolor{gray!70}0.8000 & \cellcolor{gray!70}0.5714 & & 0.7406 & \cellcolor{gray!70}0.7662 & \cellcolor{gray!70}0.7532 \\
\midrule
HGT-None & \multirow{5}{*}{Dronology} & 0.3818 & 0.6180 & 0.4710 & \multirow{5}{*}{Drools} & \cellcolor{gray!40}0.8432 & 0.8202 & 0.8310 \\
HGT-Code dependency & & \cellcolor{gray!70}0.4466 & 0.6359 & \cellcolor{gray!40}0.5231 & & \cellcolor{gray!20}0.8420 & \cellcolor{gray!40}0.9333 & \cellcolor{gray!40}0.8847 \\
HGT-User feedback & & \cellcolor{gray!40}0.4357 & \cellcolor{gray!40}0.6641 & \cellcolor{gray!70}0.5234 & & 0.8360 & \cellcolor{gray!20}0.8381 & \cellcolor{gray!20}0.8366 \\
HGT-Fine grained & & 0.4024 & \cellcolor{gray!20}0.6462 & 0.4943 & & 0.8385 & 0.8262 & 0.8317 \\
HGT-All & & \cellcolor{gray!20}0.4295 & \cellcolor{gray!70}0.6744 & \cellcolor{gray!20}0.5206 & & \cellcolor{gray!70}0.8595 & \cellcolor{gray!70}0.9441 & \cellcolor{gray!70}0.8989 \\
\midrule
HGT-None & \multirow{5}{*}{eAnci} & 0.5725 &0.8375 & 0.6792 & \multirow{5}{*}{Groovy} & 0.4168 & 0.6611 & 0.5095 \\
HGT-Code dependency & & \cellcolor{gray!40}0.6108 & 0.8679 & \cellcolor{gray!40}0.7155 & & \cellcolor{gray!40}0.4273 & 0.6462 & 0.5130 \\
HGT-User feedback & & \cellcolor{gray!20}0.6088 & \cellcolor{gray!20}0.8714 & \cellcolor{gray!20}0.7144 & & \cellcolor{gray!70}0.4442 & \cellcolor{gray!20}0.6833 & \cellcolor{gray!40}0.5342 \\
HGT-Fine grained & & 0.5835 & \cellcolor{gray!40}0.8786 & 0.7004 & & 0.4087 & \cellcolor{gray!40}0.7056 & \cellcolor{gray!20}0.5166 \\
HGT-All & & \cellcolor{gray!70}0.6438 & \cellcolor{gray!70}0.9268 & \cellcolor{gray!70}0.7590 & & \cellcolor{gray!20}0.4242 & \cellcolor{gray!70}0.7778 & \cellcolor{gray!70}0.5490 \\
\midrule
HGT-None & \multirow{5}{*}{Infinispan} & \cellcolor{gray!70}0.8742 & 0.8478 & 0.8605 & \multirow{5}{*}{iTrust} & 0.4905 & 0.6160 & 0.5440 \\
HGT-Code dependency & & 0.8551 & \cellcolor{gray!40}0.9730 & \cellcolor{gray!40}0.9101 & & \cellcolor{gray!40}0.6245 & \cellcolor{gray!70}0.8280 & \cellcolor{gray!40}0.7089 \\
HGT-User feedback & & \cellcolor{gray!40}0.8647 & 0.8613 & \cellcolor{gray!20}0.8626 & & \cellcolor{gray!20}0.5677 & 0.6520 & \cellcolor{gray!20}0.6020 \\
HGT-Fine grained & & \cellcolor{gray!20}0.8575 & \cellcolor{gray!20}0.8676 & 0.8621 & & 0.5194 & \cellcolor{gray!20}0.6640 & 0.5807 \\
HGT-All & & 0.8539 & \cellcolor{gray!70}0.9748 & \cellcolor{gray!70}0.9102 & & \cellcolor{gray!70}0.6352 & \cellcolor{gray!40}0.8200 & \cellcolor{gray!70}0.7124 \\
\midrule
HGT-None & \multirow{5}{*}{maven} & \cellcolor{gray!40}0.6545 & 0.5800 & 0.6085 & \multirow{5}{*}{Pig} & 0.5689 & 0.6815 & 0.6188 \\
HGT-Code dependency & & 0.6191 & \cellcolor{gray!70}0.8667 & \cellcolor{gray!40}0.7222 & & \cellcolor{gray!70}0.6462 & \cellcolor{gray!40}0.7778 & \cellcolor{gray!40}0.7059 \\
HGT-User feedback & & 0.5832 & 0.5800 & 0.5666 & & \cellcolor{gray!40}0.6308 & \cellcolor{gray!20}0.7593 & \cellcolor{gray!20}0.6891 \\
HGT-Fine grained & & \cellcolor{gray!20}0.6404 & \cellcolor{gray!20}0.6267 & \cellcolor{gray!20}0.6307 & & 0.5647 & 0.7056 & 0.6253 \\
HGT-All & & \cellcolor{gray!70}0.7168 & \cellcolor{gray!40}0.8200 & \cellcolor{gray!70}0.7620 & & \cellcolor{gray!20}0.6241 & \cellcolor{gray!70}0.8704 & \cellcolor{gray!70}0.7258 \\
\midrule
HGT-None & \multirow{5}{*}{Seam2} & \cellcolor{gray!40}0.7488 & 0.7370 & 0.7409 & \multirow{5}{*}{smos} & 0.5065 & 0.8990 & 0.6466 \\
HGT-Code dependency & & \cellcolor{gray!20}0.7472 & \cellcolor{gray!70}0.8870 & \cellcolor{gray!40}0.8087 & & 0.5050 & 0.8765 & 0.6405 \\
HGT-User feedback & & 0.6771 & 0.6935 & 0.6815 & & \cellcolor{gray!20}0.5083 & \cellcolor{gray!20}0.9020 & \cellcolor{gray!20}0.6502 \\
HGT-Fine grained & & 0.7217 & \cellcolor{gray!20}0.7739 & \cellcolor{gray!20}0.7447 & & \cellcolor{gray!40}0.5217 & \cellcolor{gray!40}0.9412 & \cellcolor{gray!40}0.6713 \\
HGT-All & & \cellcolor{gray!70}0.7911 & \cellcolor{gray!40}0.8696 & \cellcolor{gray!70}0.8244 & & \cellcolor{gray!70}0.5243 & \cellcolor{gray!70}0.9510 & \cellcolor{gray!70}0.6760 \\
\bottomrule
\end{tabular}
}
\end{table}

\begin{table}[htbp]
\centering
\caption{Ablation results of strategy in Gemini on the Requirements-to-Code Traceability Link Recovery Task (Three gray background colors, from light to dark, are used to highlight the top three results (Third, Second, First))}
\label{gemini_ablation}
{\footnotesize
\begin{tabular}{llccclccc}
\toprule
Approach & Project & Precision & Recall & F1-Score & Project & Precision & Recall & F1-Score \\
\midrule
Gemini-None & \multirow{5}{*}{Albergate} & \cellcolor{gray!20}0.8000 & \cellcolor{gray!20}0.6667 & \cellcolor{gray!20}0.7273 & \multirow{5}{*}{iTrust} & \cellcolor{gray!20}0.8696 & 0.7407 & 0.8000 \\
Gemini-Code dependency &  & 0.6667 & \cellcolor{gray!70}1.0000 & \cellcolor{gray!40}0.8000 &  & \cellcolor{gray!40}0.8800 & \cellcolor{gray!20}0.8148 & \cellcolor{gray!40}0.8462 \\
Gemini-User feedback &  & \cellcolor{gray!20}0.8000 & \cellcolor{gray!20}0.6667 & \cellcolor{gray!20}0.7273 &  & 0.7072 & \cellcolor{gray!40}0.9630 & \cellcolor{gray!20}0.8125 \\
Gemini-Fine grained & & \cellcolor{gray!70}1.0000 & 0.5000 & 0.6667 &  & 0.7333 & 0.7407 & 0.7370 \\
Gemini-All &  & \cellcolor{gray!40}0.8333 & \cellcolor{gray!40}0.8333 & \cellcolor{gray!70}0.8333 & & \cellcolor{gray!70}0.9000 & \cellcolor{gray!70}1.0000 & \cellcolor{gray!70}0.9474 \\
\midrule
Gemini-None & \multirow{5}{*}{eAnci} & \cellcolor{gray!20}0.5490 & \cellcolor{gray!20}0.4912 & \cellcolor{gray!20}0.5185 & \multirow{5}{*}{Dronology} & 0.5745 & \cellcolor{gray!70}0.6000 & 0.5870 \\
Gemini-Code dependency &  & 0.5417 & \cellcolor{gray!70}0.6209 & \cellcolor{gray!40}0.5786 & & \cellcolor{gray!40}0.8750 & \cellcolor{gray!20}0.4667 & \cellcolor{gray!40}0.6087 \\
Gemini-User feedback &  & \cellcolor{gray!40}0.6111 & \cellcolor{gray!40}0.5135 & \cellcolor{gray!20}0.5581 &  & \cellcolor{gray!20}0.8696 & 0.4444 & \cellcolor{gray!20}0.5882 \\
Gemini-Fine grained & & \cellcolor{gray!20}0.5490 & \cellcolor{gray!20}0.4912 & 0.5185 &  & \cellcolor{gray!70}0.8889 & 0.3556 & 0.5079 \\
Gemini-All & & \cellcolor{gray!70}0.7714 & 0.4737 & \cellcolor{gray!70}0.5870 & & 0.6585 & \cellcolor{gray!70}0.6000 & \cellcolor{gray!70}0.6279 \\
\midrule
Gemini-None & \multirow{5}{*}{Groovy} & \cellcolor{gray!20}0.8785 & 0.6667 & 0.7581 & \multirow{5}{*}{Infinispan} & \cellcolor{gray!70}0.9111 & 0.5857 & 0.7130 \\
Gemini-Code dependency & & 0.7143 & \cellcolor{gray!70}0.8333 & \cellcolor{gray!20}0.7692 &  & 0.8765 & \cellcolor{gray!40}0.6750 & \cellcolor{gray!70}0.7627 \\
Gemini-User feedback &  & \cellcolor{gray!70}0.8824 & \cellcolor{gray!70}0.8333 & \cellcolor{gray!70}0.8571 & & \cellcolor{gray!40}0.9031 & \cellcolor{gray!20}0.6250 & \cellcolor{gray!20}0.7387 \\
Gemini-Fine grained & & 0.7778 & 0.6667 & 0.7180 & & \cellcolor{gray!40}0.9031 & 0.5688 & 0.6980 \\
Gemini-All & & \cellcolor{gray!70}0.8824 & \cellcolor{gray!70}0.8333 & \cellcolor{gray!70}0.8571 & & 0.8435 & \cellcolor{gray!70}0.6929 & \cellcolor{gray!40}0.7608 \\
\midrule
Gemini-None & \multirow{5}{*}{Derby} & \cellcolor{gray!40}0.9362 & 0.6217 & 0.7472 & \multirow{5}{*}{Drools} & \cellcolor{gray!70}0.8333 & 0.5275 & 0.6460 \\
Gemini-Code dependency &  & 0.9080 & \cellcolor{gray!40}0.6991 & \cellcolor{gray!40}0.7900 &  & 0.7123 & \cellcolor{gray!70}0.6047 & \cellcolor{gray!20}0.6541 \\
Gemini-User feedback & & \cellcolor{gray!40}0.9362 & \cellcolor{gray!20}0.6525 & \cellcolor{gray!20}0.7690 &  & \cellcolor{gray!70}0.8333 & \cellcolor{gray!20}0.5425 & \cellcolor{gray!40}0.6572 \\
Gemini-Fine grained &  & \cellcolor{gray!70}0.9412 & 0.5125 & 0.6636 & & 0.7119 & 0.4884 & 0.5793 \\
Gemini-All &  & 0.7557 & \cellcolor{gray!70}0.8761 & \cellcolor{gray!70}0.8115 &  & \cellcolor{gray!20}0.8065 & \cellcolor{gray!40}0.5814 & \cellcolor{gray!70}0.6757 \\
\midrule
Gemini-None & \multirow{5}{*}{maven} & 0.7000 & \cellcolor{gray!20}0.7778 & 0.7368 & \multirow{5}{*}{Pig} & \cellcolor{gray!20}0.6727 & 0.6271 & 0.6491 \\
Gemini-Code dependency &  & \cellcolor{gray!40}0.8125 & 0.7222 & \cellcolor{gray!20}0.7647 & & \cellcolor{gray!40}0.6825 & \cellcolor{gray!20}0.6525 & \cellcolor{gray!20}0.6672 \\
Gemini-User feedback &  & \cellcolor{gray!70}0.8333 & \cellcolor{gray!40}0.8333 & \cellcolor{gray!40}0.8333 &  & 0.6398 & \cellcolor{gray!40}0.7015 & \cellcolor{gray!40}0.6692 \\
Gemini-Fine grained &  & 0.6667 & \cellcolor{gray!20}0.7778 & 0.7179 &  & \cellcolor{gray!70}0.8929 & 0.4237 & 0.5747 \\
Gemini-All &  & \cellcolor{gray!20}0.8095 & \cellcolor{gray!70}0.9444 & \cellcolor{gray!70}0.8718 & & 0.6575 & \cellcolor{gray!70}0.8136 & \cellcolor{gray!70}0.7273 \\
\midrule
Gemini-None & \multirow{5}{*}{Seam2} & \cellcolor{gray!20}0.6538 & 0.5235 & 0.5814 & \multirow{5}{*}{smos} & \cellcolor{gray!70}0.7714 & 0.2500 & 0.3776 \\
Gemini-Code dependency & & \cellcolor{gray!70}0.6857 & \cellcolor{gray!20}0.6000 & \cellcolor{gray!20}0.6400 &  & 0.5303 & \cellcolor{gray!20}0.3241 & \cellcolor{gray!20}0.4023 \\
Gemini-User feedback &  & \cellcolor{gray!40}0.6667 & \cellcolor{gray!40}0.6500 & \cellcolor{gray!40}0.6582 &  & \cellcolor{gray!20}0.5623 & \cellcolor{gray!40}0.3254 & \cellcolor{gray!40}0.4122 \\
Gemini-Fine grained &  & 0.5358 & 0.5583 & 0.5483 & & \cellcolor{gray!40}0.6857 & 0.2222 & 0.3357 \\
Gemini-All & & 0.6222 & \cellcolor{gray!70}0.7000 & \cellcolor{gray!70}0.6588 &  & 0.5556 & \cellcolor{gray!70}0.3704 & \cellcolor{gray!70}0.4444 \\
\bottomrule
\end{tabular}
}
\end{table}

Among these strategies, code dependency relations yield the highest performance improvement, followed by user feedback and fine-grained associations. This suggests that code dependency information is particularly effective in revealing inter-code relationships and guiding the model to associate relevant code artifacts with a given requirement, ultimately improving the accuracy of trace link prediction. Additionally, the abundance of code dependency edges enhances message passing and provides more structural information for HGT to leverage.

Moreover, the HGT-All variant, which integrates all three strategies, achieves the highest average F1-score of 0.7219 and significantly outperforms other variant and HGT-None at the 0.05 confidence level across all evaluated projects. This demonstrates that HGT can effectively integrate heterogeneous strategy information, and its attention mechanism is capable of distinguishing the relative importance of each edge type. Consequently, HGT-All benefits from a more precise and discriminative use of structural context, further enhancing its predictive capabilities.

For Gemini, we observe that Gemini-All (average F1 = 0.7336) achieves the best overall performance, followed by Gemini-Code Dependency (average F1 = 0.6903), Gemini-User Feedback  (average F1 = 0.6901), and Gemini-None (average F1 = 0.6535), while Gemini-Fine Grained (average F1 = 0.6055) performs the worst. These results indicate that incorporating multiple types of strategy edges into the model can effectively maximize performance gains.

From the perspective of strategy usage, both Gemini-Code Dependency and Gemini-User Feedback outperform Gemini-None, which does not use any strategy edges. We believe this is because the Code Dependency and User Feedback strategies provide high-quality supplementary information from the perspectives of code structure and real trace links, respectively, thus contributing to improved model performance.

Meanwhile, we find that the Fine-Grained strategy performs the worst. This may be due to the introduction of a large amount of noisy information during the construction of this strategy. Although it contains some valid strategy edges, the absence of ground-truth supervision prevents Gemini from distinguishing effective edges from noisy or misleading ones, resulting in a significant performance drop. This further indicates that, under zero-shot settings, the model is highly sensitive to the accuracy of strategy edges. Noisy data has a particularly pronounced negative impact, especially when compared to supervised models like HGT, which can learn to identify and filter useful edges through training.

\begin{tcolorbox}[title=Observation 5]
Both HGT and Gemini can effectively integrate multiple strategies. HGT-All and Gemini-All outperform their respective base models, HGT-None and Gemini-None, in terms of F1 performance.
\end{tcolorbox}

\subsubsection{Answer RQ5: How do HGT-All and Gemini-All perform in the requirements-to-code traceability link recovery task?}

\begin{table}[htbp]
\caption{Performance comparison of HGT-All, Gemini-All, and baselines on the Requirements-to-Code Traceability Link Recovery Task (Three gray background colors, from light to dark, are used to highlight the top three results (Third, Second, First))}
\label{performance_comparision}
\centering
{\footnotesize
\begin{tabular}{llccclccc}
\toprule
Approach & Project & Precision & Recall & F1-Score & Project & Precision & Recall & F1-Score \\
\midrule
TAROT   &\multirow{5}{*}{Albergate}   &  0.2785   &0.3510   & 0.3106&\multirow{5}{*}{Infinispan} & 0.2378 & 0.1577 & 0.1897\\
GA-XWCoDe   &&  \cellcolor{gray!40}0.5000   &\cellcolor{gray!20}0.5000 &\cellcolor{gray!40}0.5000&& 0.6842 & 0.6309 & 0.6523  \\
HGNNLink&   & \cellcolor{gray!40} 0.5000 & 0.3333  & 0.4000 && \cellcolor{gray!70}0.8733 & \cellcolor{gray!40}0.9631 & \cellcolor{gray!70}0.9155\\
HGT-All&  & 0.4189 & \cellcolor{gray!40}0.6200 & \cellcolor{gray!20}0.4914&& \cellcolor{gray!40}0.8539 & \cellcolor{gray!70}0.9748 & \cellcolor{gray!40}0.9102 \\
Gemini-All&  & \cellcolor{gray!70}0.8333 & \cellcolor{gray!70}0.8333 & \cellcolor{gray!70}0.8333&& \cellcolor{gray!20}0.8435 & \cellcolor{gray!20}0.6929 & \cellcolor{gray!20}0.7608\\
\midrule
TAROT &\multirow{5}{*}{Derby} &  0.3006 & 0.3854 & 0.3378&\multirow{5}{*}{iTrust} & 0.5750 & 0.4510 & 0.5055\\
GA-XWCoDe &&  0.6385 & 0.6052 & 0.6192&& \cellcolor{gray!40}0.7698 & \cellcolor{gray!40}0.7597 & 0.7619 \\
HGNNLink &&  \cellcolor{gray!20}0.7378 & \cellcolor{gray!20}0.7299 & \cellcolor{gray!20}0.7298&& 0.6097 & 0.7936\cellcolor{gray!20} & 0.6861\\
HGT-All &&  \cellcolor{gray!40}0.7406 & \cellcolor{gray!40}0.7662 & \cellcolor{gray!40}0.7532&& 0.6352\cellcolor{gray!20} &\cellcolor{gray!40} 0.8200 & 0.7124\cellcolor{gray!20} \\
Gemini-All &&  \cellcolor{gray!70}0.7557 & \cellcolor{gray!70}0.8761 & \cellcolor{gray!70}0.8115  && \cellcolor{gray!70}0.9000 & \cellcolor{gray!70}1.0000 & \cellcolor{gray!70}0.9474\\
\midrule
TAROT &\multirow{5}{*}{Dronology} &  0.1785 & 0.4088 & 0.2485&\multirow{5}{*}{maven} & 0.2881 & 0.3377 & 0.3110\\
GA-XWCoDe &&  0.3777 & 0.5875 & 0.4598&& 0.6306 & 0.5914 & 0.6038\\
HGNNLink &&  \cellcolor{gray!20}0.4198 & \cellcolor{gray!40}0.6123 & \cellcolor{gray!20}0.4960&&\cellcolor{gray!40} 0.7358 & \cellcolor{gray!20}0.8080 & \cellcolor{gray!40}0.7648 \\
HGT-All &&  \cellcolor{gray!40}0.4295 & \cellcolor{gray!70}0.6744 & \cellcolor{gray!40}0.5206&& \cellcolor{gray!20}0.7168 &\cellcolor{gray!40} 0.8200 & \cellcolor{gray!20}0.7620\\
Gemini-All &&  \cellcolor{gray!70}0.6585 & \cellcolor{gray!20}0.6000 & \cellcolor{gray!70}0.6279 && \cellcolor{gray!70}0.8095 & \cellcolor{gray!70}0.9444 &\cellcolor{gray!70} 0.8718 \\
\midrule
TAROT &\multirow{5}{*}{Drools} &  0.1610 & 0.1736 & 0.1670& \multirow{5}{*}{Pig} & 0.2948 & 0.2806 & 0.2822\\
GA-XWCoDe &&  0.6263 & 0.5589 & 0.5797&& \cellcolor{gray!70}0.7423 & 0.6589 & 0.6898 \\
HGNNLink &&  \cellcolor{gray!70}0.8632 & \cellcolor{gray!40}0.9336 & \cellcolor{gray!40}0.8961&& \cellcolor{gray!20}0.6568 & \cellcolor{gray!70}0.9196 & \cellcolor{gray!70}0.7644 \\
HGT-All & & \cellcolor{gray!40}0.8595 & \cellcolor{gray!70}0.9441 & \cellcolor{gray!70}0.8989&& 0.6241 & \cellcolor{gray!40}0.8704 & \cellcolor{gray!20}0.7258 \\
Gemini-All &&  \cellcolor{gray!20}0.8065 & \cellcolor{gray!20}0.5814 & \cellcolor{gray!20}0.6757 && \cellcolor{gray!40}0.6575 & \cellcolor{gray!20}0.8136 & \cellcolor{gray!40}0.7273 \\
\midrule
TAROT &\multirow{5}{*}{eAnci} &  0.2351& 0.3155&0.2694 &\multirow{5}{*}{Seam2} & 0.2892 & 0.3542 & 0.3184\\
GA-XWCoDe &&  \cellcolor{gray!20}0.6215 & \cellcolor{gray!20}0.5325 & 0.5736&& \cellcolor{gray!70}0.7423 & 0.6589 & 0.6898 \\
HGNNLink &&  0.6154 & \cellcolor{gray!40}0.8896 & \cellcolor{gray!40}0.7275&& \cellcolor{gray!20}0.6568 &\cellcolor{gray!70} 0.9196 & \cellcolor{gray!70}0.7644\\
HGT-All &&  \cellcolor{gray!40}0.6438 & \cellcolor{gray!70}0.9268 & \cellcolor{gray!70}0.7590&& 0.6241 & \cellcolor{gray!40}0.8704 & \cellcolor{gray!20}0.7258 \\
Gemini-All &&  \cellcolor{gray!70}0.7714 & 0.4737 & \cellcolor{gray!20}0.5870&& \cellcolor{gray!40}0.6575 & \cellcolor{gray!20}0.8136 & \cellcolor{gray!40}0.7273 \\
\midrule
TAROT &\multirow{5}{*}{Groovy} &  0.3558 & 0.4111 & 0.3814&\multirow{5}{*}{smos} & 0.2892 & 0.5158 & 0.3706 \\
GA-XWCoDe &&  0.3577 & 0.6133 & 0.4598&& \cellcolor{gray!70}0.8910 & \cellcolor{gray!40}0.8703 & \cellcolor{gray!70}0.8797 \\
HGNNLink && \cellcolor{gray!20}0.3768 & \cellcolor{gray!20}0.6500 & \cellcolor{gray!20}0.4749&& 0.5031 & \cellcolor{gray!20}0.8371 & \cellcolor{gray!20}0.6194\\
 HGT-All && \cellcolor{gray!40}0.4242 & \cellcolor{gray!40}0.7778 & \cellcolor{gray!40}0.5490&& \cellcolor{gray!20}0.5243 & \cellcolor{gray!70}0.9510 & \cellcolor{gray!40}0.6760\\
Gemini-All &&  \cellcolor{gray!70}0.8824 & \cellcolor{gray!70}0.8333 & \cellcolor{gray!70}0.8571&& \cellcolor{gray!40}0.5556 & 0.3704 & 0.4444 \\

\bottomrule
\end{tabular}
}
\label{tab:example5col}
\end{table}

The experimental results for this RQ are presented in Table \ref{performance_comparision}. For supervised methods, HGT-All shows a significant advantage over the two supervised baseline models, GA-XWCoDe and HGNNLink, at a confidence level of $\alpha < 0.05$. GA-XWCoDe is a model that performs trace link prediction based on XGBoost, with parameters optimized through a genetic algorithm. It incorporates structural dependency information from code artifacts and assigns weights to different dependency relations, allowing them to influence XGBoost’s link generation decisions with varying degrees of importance. This model demonstrates strong generalization capabilities. However, it primarily relies on traditional machine learning techniques, and the effectiveness of its feature construction limits its performance. Additionally, it lacks the ability to deeply model user feedback and fine-grained semantic information. In contrast, HGT-All introduces a heterogeneous graph learning mechanism that integrates strategies such as code dependency relationships, user feedback, and fine-grained textual similarity between requirements and code, enabling more comprehensive strategy fusion. Compared to HGNNLink, HGT-All also incorporates optimizations from the perspective of strategy integration, further enhancing the dimensionality of information represented in the heterogeneous graph.

In the comparison of unsupervised methods, Gemini-All significantly outperforms the information retrieval-based method TAROT at the $\alpha<0.05$ confidence level. TAROT is built on traditional IR models such as the Vector Space Model (VSM) and Latent Semantic Indexing (LSI). It relies on term frequency statistics and shallow semantic similarity, which limits its ability to capture deep semantic associations between requirement documents and code implementations. In contrast, Gemini 2.5 pro has been trained on a large-scale corpus and possesses stronger semantic representation capabilities, enabling it to more accurately identify potential links between requirements and code even without labeled data, through contextual understanding and optimized code representations.

Furthermore, we observed that Gemini-All delivers more pronounced performance improvements on small-scale projects. For instance, in \textit{Albergate}, the smallest of the 12 evaluated projects, Gemini-All achieves an F1-score that is 69.58\% higher than HGT-All. Conversely, in large-scale industrial projects, HGT-All tends to be more advantageous; for example, in \textit{Infinispan,} the largest among the 12 projects, HGT-All outperforms Gemini-All by 19.64\% in F1-score. We hypothesize that this is because large-scale projects contain more complex structural relationships, which can enrich the heterogeneous graph with more meaningful edge relations, thereby enhancing the model’s learning capacity. On the other hand, in small-scale projects, the semantic descriptions of code artifacts tend to be closer to the requirement artifacts, as these projects are more likely to employ direct implementation approaches that do not depend on complex intermediate abstraction layers (e.g., Data Access Object (DAO) layers in JavaWeb development).

Considering the computational cost and resource demands of both HGT-All and Gemini-All, we recommend the following: For large-scale projects where trace link generation time is not a strict constraint, HGT-All is preferable; for small-scale projects with sufficient budget for computational resources, Gemini-All is more suitable.

\begin{tcolorbox}[title=Observation 6]
Both HGT-All and Gemini-All demonstrate strong performance advantages in the requirements-to-code traceability link recovery (TLR) task. At a confidence level of $\alpha < 0.05$, HGT-All and Gemini-All significantly outperform all other baseline methods.
\end{tcolorbox}

\section{Discussion} \label{sec discussion}
This section first interprets the results of this study and then discusses the implication of the study results for researchers and practitioners. 

\subsection{Interpretation of Results}
\textbf{Interpretation of RQ1 to RQ3 Results}: In the experiments of RQ1 to RQ3, we validated the core issue raised in the introduction: in NL-PL TLR tasks, relying solely on textual similarity is insufficient to establish high-quality links. It is essential to incorporate richer auxiliary strategies to achieve more accurate and robust link prediction. Based on our observation and analysis of the experimental results, we believe this phenomenon stems from the significant semantic gap between NL and PL artifacts.

First, PL artifacts are often semantically abstract and structurally complex. Particularly in cases where comments are sparse or lack informative content, semantically rich elements such as class names and method names occupy only a small portion of the code. This makes it difficult to accurately convey functional semantics. Additionally, the presence of numerous language-specific keywords and complex API calls further increases the difficulty for language models to understand the semantics of PL artifacts, thereby widening the semantic gap between NL and PL. To address this challenge, the Fine-Grained strategy examined in RQ3 decomposes PL artifacts into smaller semantic units (such as class names, method names, comments, parameter types, return types, etc.) and matches them separately with the NL artifact. This significantly enhances the model’s ability to capture local semantic information. Even when overall textual similarity is low, effective traceability links can still be identified through local alignments, thus alleviating semantic mismatch.

Second, due to the modular nature of PL artifacts, a single NL artifact often corresponds to multiple PL artifacts. Some of the implementation details may be only briefly mentioned in the NL artifact, further limiting the effectiveness of textual similarity-based methods. To address this, two auxiliary strategies proposed in RQ3 — Traceability Link Propagation for Homogeneous Artifacts and Code Dependency — compensate for the inability of textual similarity approaches to leverage structural features. These strategies enhance cross-artifact-type link recovery by propagating similarity among homogeneous artifacts and modeling modular invocation relationships.

Moreover, the results of RQ2 indicate that textual similarity-based methods perform best in the Test–Code TLR task, while their performance is poorest in the Requirements–Code task. We attribute this to the fact that test code typically adopts naming conventions similar or derived from the source code under test, resulting in high textual consistency. In contrast, test cases often directly reference the implementation code rather than the functional requirements described in natural language, leading to significant differences in expression between requirements and tests. Therefore, textual similarity performs poorly in this context. To mitigate this issue, the Leveraging Intermediate Artifacts strategy introduced in RQ3 proposes using source code as an intermediate artifact to establish indirect links between requirements and tests, thereby reducing the semantic gap between them.

\textbf{Interpretation of RQ4 to RQ5 Results}: All of the above observations support the conclusion that textual similarity alone is inadequate for NL-PL TLR tasks, and that auxiliary strategies can significantly enhance the performance of such methods. However, it is noteworthy that although various auxiliary strategies have been proposed, existing research typically employs them in isolation, with limited attempts at multi-strategy integration. The experimental results of RQ4 demonstrate that our proposed methods - HGT-All and Gemini-All - can effectively integrate multiple strategies from the perspectives of structural information modeling and language understanding, achieving better performance across different tasks compared to any single-strategy approach. Furthermore, the results of RQ5 confirm that both HGT-All and Gemini-All perform well under both supervised and unsupervised learning settings. We argue that this is due to the HGT model’s ability to represent different auxiliary strategies through multiple edge types, allowing high-quality strategy information to enhance graph representation from multiple dimensions. It also enables the propagation of information across nodes through strategy-specific edges, improving link prediction accuracy. Meanwhile, the Gemini model incorporates auxiliary information in the form of prompts as additional input, providing LLMs with a more comprehensive basis for judgment. This allows the model to go beyond surface-level textual cues and make decisions by integrating structural, semantic, and contextual information.

\subsection{Implications}

\subsubsection{Implications for Researchers}
\textbf{Integration of new strategies}: When researchers propose new auxiliary strategies, they may consider integrating them with existing ones. HGT-All and Gemini-All can serve as effective carriers for such integration, enabling synergistic effects among strategies and fully leveraging the performance potential of the newly proposed methods.

\textbf{Gemini-All relies on high-quality auxiliary information}: As a large-scale pre-trained language model without fine-tuning on specific project data, Gemini-All heavily depends on the relevance of auxiliary strategy information. If such information is weakly correlated with the actual traceability links, the model may fail to correctly interpret its significance or even be misled, ultimately degrading prediction performance. This highlights the model’s sensitivity to the semantic relevance of additional input, underscoring the need for careful design of strategy inputs in practical applications.

\textbf{HGT-All is sensitive to data scale}: The message-passing mechanism of HGT relies on rich relationships among heterogeneous nodes and edges within the graph. In small-scale projects or sparsely connected graphs, the available information for learning is limited, reducing the model’s ability to generalize. Consequently, HGT-All may experience significant performance drops under small or insufficient datasets, limiting its structural advantages.

\textbf{HGT-All requires supervised learning support}: As a GNN-based approach, HGT-All requires training on labeled data (i.e., existing ground-truth traceability links) to learn effective representations and predictive capabilities. Therefore, it cannot be directly applied to projects without annotated data. This limits its applicability in cold-start scenarios and stands in contrast to the zero-shot inference capabilities of Gemini-All.

\subsubsection{Implications for Practitioners}
\textbf{Gemini-All entails high computational and financial costs}: Since Gemini-All involves querying large language model APIs, its inference process is resource-intensive and often billed based on token usage. When applied to large-scale requirement sets, this may incur substantial financial costs. Moreover, its relatively slow inference speed makes it less suitable for real-time or time-sensitive industrial scenarios. As a result, in resource-constrained or budget-limited settings, its scalability and practicality may be limited.

\textbf{HGT-All relies on high-quality labeled training data}: The performance of HGT-All is highly dependent on the quality and quantity of training data annotations. For enterprises, this places high demands on the expertise of data labeling personnel. Furthermore, meeting the model’s training requirements often necessitates significant human resource investment, leading to increased operational costs.

\textbf{A more adaptable replacement with suitable models}: Enterprise developers can replace the HGT and Gemini 2.5 Pro models used in this study with internally fine-tuned heterogeneous graph neural networks or LLMs to better accommodate project-specific needs. As more advanced models continue to emerge, the approach proposed in this paper can serve as a general paradigm, enabling flexible adaptation to newer models and further improving overall performance, thereby enhancing its practical potential in industry applications.

\section{Threats to Validity} \label{tv}
To comprehensively assess the credibility and applicability of this study, we systematically discuss potential threats to validity from four dimensions: internal validity, external validity, construct validity, and conclusion validity.

\textbf{Internal Validity}: A potential threat to internal validity arises from the presence of Java Servlet Page (JSP) files in the iTrust project. Since the Spoon package used in this study cannot analyze the code structure within JSP files, it fails to extract code dependency relationships from these files. This limitation may affect the completeness of structural information used by the model in this project, potentially influencing the experimental results. However, JSP pages constitute only a small portion of the iTrust project, so the overall impact is limited. To further improve internal validity, future work could explore integrating static analysis tools capable of handling JSP files.
 
\textbf{External Validity}: To enhance the generalizability of our findings, this study evaluates the proposed methods on 12 open-source Java projects of varying domains and sizes. This helps reduce the threat to external validity by covering a broader range of application scenarios. Nevertheless, the current approach relies on Java-based static analysis tools for extracting code dependencies and does not support projects in other programming languages, such as Python or C++. This limitation restricts the applicability of the method to multi-language software projects. Therefore, future work will consider extending support to additional languages to improve the model's generalizability and adaptability.
 
\textbf{Construct Validity}: The performance evaluation in this study employs three widely accepted metrics — Precision, Recall, and F1-score. These metrics have been shown to effectively measure the accuracy and stability of TLR systems \cite{wang2024empirical, wang2022systematic}, thereby supporting the construct validity of the evaluation.
  
\textbf{Conclusion Validity}: To ensure that the reported performance improvements are statistically significant, we adopt non-parametric hypothesis testing methods, including Spearman’s rank correlation and the Wilcoxon signed-rank test. These statistical methods help reduce the influence of random variation, thereby mitigating threats to conclusion validity.

\section{Related Work}\label{related work}
In this section, we present the related work of this study from three perspectives: empirical research on ST, the application of graph neural networks in ST, and the application of large language models in ST.

\subsection{Empirical Research on Software Traceability}
At present, several empirical studies in the ST field have guided practitioners and researchers from different perspectives to improve their work. For example, Wang et al. \cite{wang2024empirical} conducted an empirical study on SOTA methods in the task of recovering requirements-to-code traceability links, and called for researchers to adopt standardized evaluation metrics and baselines to enhance the credibility of research results. Wang et al. \cite{wang2023empirical} also investigated the issue of data imbalance in machine learning-based TLR tasks and found that data balancing techniques have a positive impact on the performance of ML-based TLR. Ali et al. \cite{ali2015empirical} used eye-tracking devices to statistically analyze the time software developers spent focusing on different parts of code artifacts while reading code. The empirical results showed that developers pay more attention to method names and comments, indicating that future research on TLR involving code artifacts should place greater emphasis on these elements. Charalampidou et al. \cite{charalampidou2021empirical} conducted an empirical study from multiple dimensions, including types of artifacts involved in TLR, the benefits TLR brings to the software engineering field, the methods used to measure those benefits, and the research methods applied. Their findings revealed that requirement artifacts dominate in TLR, and that current research mainly focuses on proposing and establishing new TLR techniques. Oliveto et al. \cite{oliveto2010equivalence} statistically analyzed the equivalence of various IR-based TLR methods, showing that combining Latent Dirichlet Allocation (LDA) with other IR techniques can improve the TLR accuracy of standalone methods. 

\subsection{Graph Neural Networks used in Software Traceability}
In the field of TLR, several researchers have introduced Graph Neural Networks (GNNs) to enhance link prediction performance. For example, Wang et al. \cite{wang2025HGNNLink} conducted a comparative performance study of Heterogeneous Graph Attention Networks (HAN) \cite{wang2019heterogeneous}, HGT, and Relational Graph Convolutional Networks (R-GCN) \cite{schlichtkrull2018modeling}, concluding that HGT demonstrated the best performance. Based on this finding, they integrated HGT with code dependency relations and IR results to perform link prediction, establishing a SOTA method in the domain. Zou et al. \cite{zou2024enhancing} applied Node2vec \cite{grover2016node2vec} to embed both code dependency matrices and artifact co-occurrence matrices, and calculated the importance of each type of code dependency using cosine similarity. These features were then combined with an XGBoost model to achieve more accurate link prediction. Bai et al. \cite{bai2024improving} proposed the AIPL method, which extracts information from four GitHub-related sources associated with pull requests (PRs)—repositories, users, issues, and PRs—and models them as task-specific heterogeneous graphs. By constructing metapaths between issues and various sources, the method enhances the accuracy of PR-Issue traceability link prediction. Wang et al. \cite{wang2019using} extracted structural information from source code to construct code relation graphs and applied graph embedding techniques to integrate this structural information with traditional IR approaches, improving the overall performance of TLR tasks.

\subsection{Large Language Models used in Software Traceability}
LLMs possess powerful semantic understanding capabilities and have been widely adopted by researchers in the ST field in recent years. For example, Wang et al. \cite{wang2025mplinker} proposed a Multi-template Prompt Learning method with adversarial training, called MPLinker, to achieve highly generalized Issue-Commit Traceability Link Recovery (TLR). Hassine et al. \cite{hassine2024llm} introduced a method based on GPT-3.5-turbo \cite{brown2020language} to generate security-related traceability links between requirements expressed in natural language and goals described as part of GRL models. Evaluation results demonstrated the effectiveness of this approach. Fuchb et al. \cite{fuchss2025lissa} employed Retrieval-Augmented Generation (RAG) to retrieve the top-k most similar artifacts and used GPT-4O \cite{achiam2023gpt} and GPT-4O-mini as generators to perform zero-shot traceability link question answering. Ali et al. \cite{ali2024establishing} attempted to combine RAG with graph-based indexing techniques, extracting structural information from artifacts via knowledge graphs to further enhance TLR performance.

\subsection{Conclusive Summary}
In summary, from the perspective of empirical research in TLR, existing literature has not systematically investigated relevant techniques from a methodological strategy standpoint. Moreover, few studies have explored the applicability and effectiveness of text similarity methods across different types of software artifacts using large-scale datasets. In contrast, this study introduces the Different Ratio metric and provides empirical evidence that text similarity methods alone perform suboptimally in TLR tasks involving NL-PL artifacts. Furthermore, we systematically analyze various auxiliary strategies proposed in previous work, highlighting their necessity and underlying mechanisms.

From the perspective of applying graph neural networks and large language models, existing approaches have yet to fully exploit and integrate multi-source strategic information. This study explores such integration through two complementary avenues: modeling edge information in heterogeneous graphs and leveraging additional task-specific descriptions in prompts. Empirical results demonstrate that combining these sources significantly enhances the performance of TLR tasks.

\section{Conclusions and Future Work} \label{conclusion}
This study systematically reviews and analyzes research developments in the field of software traceability link recovery over the past five years, with a particular focus on evaluating the effectiveness of text similarity-based methods across different types of software artifacts. The results indicate that for heterogeneous artifacts, such as in requirements-to-code TLR tasks, the performance of text similarity-based methods is suboptimal. Furthermore, we conducted a statistical analysis of the strategies employed in prior studies and proposed two models that integrate multi-source strategies: HGT-All and Gemini-All. Experimental results demonstrate that HGT-All significantly outperforms the current state-of-the-art method HGNNLink \cite{wang2025HGNNLink} in supervised settings at a confidence level of $<0.05$, while Gemini-All clearly surpasses the unsupervised state-of-the-art method TAROT\cite{gao2022using} in unsupervised settings at the same confidence level. These findings strongly validate the effectiveness of integrating diverse strategies to enhance traceability link recovery performance. This research offers new perspectives and practical approaches for combining multi-source strategies and achieving high-quality traceability link recovery for heterogeneous software artifacts.

In our future work, to enhance the interpretability and industrial applicability of our models, we plan to incorporate visualization mechanisms and explainability algorithms to present the reasoning paths and supporting evidence behind each recommended trace link, thereby improving system usability and trustworthiness for developers. Additionally, we intend to explore multi-agent frameworks to enhance task specialization and enable agents to review and correct each other’s outputs. The multi-agent approach may help mitigate the high error rates observed in certain strategies, such as fine-grained similarity analysis.


\section*{Data Availability Statements}
The replication package of this study has been made available at \cite{source_code}.

\bibliographystyle{ACM-Reference-Format}
\bibliography{acmart}

\appendix
\section{Project Details Used in the Primary Literature Collection}
The projects collected in the primary literature and their statistical details provided in Table \ref{table app A}:
\begin{longtable} {@{}p{2.5cm}p{1.3cm}p{1.3cm}p{1.3cm}p{2.2cm}p{2.8cm}@{}}
\caption{Project Details Used in the Primary Literature Collection (``SA num'' refers to the number of source artifacts, ``TA num'' refers to the number of target artifacts, ``TL'' refers to the number of true links, and ``Req'' refers to ``Requirements'').}  \label{table app A}\\
\toprule
Project & SA num& TA num& TL num& Project types  & References \\
\midrule
\endfirsthead

\multicolumn{6}{c}%
{{\bfseries \tablename\ \thetable{} -- continued from previous page}} \\
\toprule
Projects & SA num& TA num& TL num& Project types  & Reference \\
\midrule
\endhead

\midrule \multicolumn{5}{r}{{Continued on next page}} \\
\endfoot

\bottomrule
\endlastfoot

airflow & 4906 & 4614 & 5295 &  Issue-Commit&\cite{mazrae2021automated}\\
Albergate & 17 & 55 & 54 & Req-Code & \cite{li2020combining, rodriguez2020ir}\\
ambari & 23730 & 27034 & 35589 & Issue-Commit & \cite{zhang2023ealink, mazrae2021automated}\\
ant-ivy & 967 & 1073 & 1158 & Issue-Commit & \cite{lan2023btlink, zhu2024deep}\\
\multirow{2}{*}{archiva} & 1899& 5462 & 2275 & Issue-Commit & \cite{dai2023constructing}\\
&1195 & 1238 &510 & Issue-Issue & \cite{dai2023constructing}\\
arrow & 6223 & 5159 & 5252 & Issue-Commit & \cite{mazrae2021automated}\\
arthas & 122 & 167 & 167 & Issue-Commit & \cite{liu2020traceability, sun2024aviate}\\
automation club & 19 & 19 & 5 & Issue-Issue & \cite{tian2023cross}\\
Avro & 2011 & 2016 & 2542 & Issue-Commit & \cite{lan2023btlink, zhu2024deep}\\
awesome-berlin & 74 & 74 & 74 & Issue-Commit & \cite{liu2020traceability,sun2024aviate}\\
\multirow{2}{*}{Beam} & 7456 & 7488 & 2200 & Issue-Commit & \cite{zhu2024deep, lan2023btlink, mazrae2021automated}\\
& 1785 & 2107 & 644 & Issue-Issue & \cite{tian2023cross}\\
bk-cmdb & 886 & 1149 & 1179 & Issue-Commit & \cite{liu2020traceability, sun2024aviate}\\
Buildr & 532 & 479 & 829 & Issue-Commit & \cite{lan2023btlink, zhu2024deep}\\
bytedeco & 105 & 106 & 109 & Issue-Commit & \cite{majidzadeh2024multi}\\
Calcite & 2777 & 2126 & 3058 & Issue-Commit & \cite{lan2023btlink, mazrae2021automated, deng2024mtlink}\\
canal & 232 & 272 & 273 & Issue-Commit & \cite{liu2020traceability, sun2024aviate}\\
\multirow{2}{*}{Cassandra} & 267 & 145 & 146 & Issue-Commit & \cite{mazrae2021automated}\\
& 8217 & 8395 & 3363 & Issue-Issue & \cite{dai2023constructing}\\
CCHIT & 116 & 1064 & 588 &Req-Code & \cite{schlutter2020trace}\\
Cica & 25 & 27 & 27 & Issue-Commit & \cite{liu2020traceability, sun2024aviate}\\
CM1-NASA & 22 & 53 & 46 & Req-Design & \cite{wang2021analyzing, schlutter2020trace, li2024mltracer}\\
commons-io & 35 & 28 & 41 & Test-Code & \cite{sun2024method, white2022tctracer}\\
Commons Lang & 78 & 42 & 78 & Test-Code & \cite{sun2024method, white2022tctracer}\\
Derby & 390 & 2200 & 2312 & Req-Code & \cite{gao2022propagating, gao2022using, dong2022semi}\\ 
\multirow{2}{*}{Derby}& 6812 & 8061 & 7263 & Issue-Commit & \cite{dai2023constructing}\\
& 5865 & 5959 & 3529 & Issue-Issue & \cite{dai2023constructing}\\
\multirow{3}{*}{Dronology} & 28 & 184 & 393 & Req-Code & \cite{rodriguez2021leveraging, gao2024triad}\\
& 58 & 144 & 133 & Req-Design & \cite{gao2024triad}\\
& 144 & 184 & 285 & Design-Code & \cite{gao2024triad}\\
\multirow{3}{*}{Drools} & 183 & 3527 & 842 & Req-Code & \cite{gao2022propagating,gao2022using,dong2022semi}\\
& 4973 & 9256 & 5528 & Issue-Commit & \cite{dai2023constructing}\\
& 1891 & 1876 & 663 & Issue-Issue & \cite{dai2023constructing}\\
druid & 980 & 1161 & 1161 & Issue-Commit & \cite{sun2024aviate,liu2020traceability}\\
eAnci & 139 & 55 & 567 & Req-Code & \cite{peng2023enhancing, li2020combining, du2020automatic, chen2021self, zou2024xwcode, hey2021improving, zou2024enhancing}\\
\multirow{4}{*}{EasyClinic} & 30 & 47 & 93 & Req-Code & \cite{rodriguez2021leveraging, gao2024triad, li2024mltracer, wang2023df4rt}\\
& 30 & 63 & 63 & Req-Test & \cite{wang2021analyzing, wang2020automated, du2020automatic, li2024mltracer}\\
& 63 & 47 & 204 & Test-Code & \cite{du2020automatic}\\
& 20 & 63 & 83 & Design-Test & \cite{du2020automatic}\\
\multirow{2}{*}{EBT}& 41 & 50 & 98 & Req-Code&\cite{moran2020improving, rodriguez2021leveraging, gao2024triad, rodriguez2020ir}\\
& 41 & 25 & 51 & Req-Test& \cite{wang2020automated}\\
Emmagee & 31 & 32 & 32 & Issue-Commit & \cite{liu2020traceability, sun2024aviate}\\
\multirow{2}{*}{errai} & 969 & 1981 & 638 & Issue-Commit & \cite{dai2023constructing}\\
& 137 & 137 & 40 & Issue-Issue & \cite{dai2023constructing}\\
eTour & 58 & 116 & 308 & Req-Code & \cite{moran2020improving, wang2020automated, fuchss2025lissa, peng2023enhancing, li2020combining, du2020automatic, chen2021self, zhang2021recovering, rodriguez2020ir, zou2024xwcode, li2024mltracer, shen2021supporting, wang2023df4rt, hey2021improving, zou2024enhancing}\\
feedHenry & 442 & 521 & 176 & Issue-Issue & \cite{tian2023cross}\\
\multirow{2}{*}{flink}& 4441 & 4881 & 1895 & Issue-Issue & \cite{dai2023constructing}\\
& 12181 & 11613 & 14472& Issue-Commit & \cite{dai2023constructing, mazrae2021automated}\\
Flask & 705 & 752 & 752 & Issue-Commit & \cite{zhu2022enhancing, zhu2024deep, lin2021traceability}\\
freemarker & 63& 172 & 177 & Issue-Commit & \cite{mazrae2021automated}\\
Gannt & 17 & 69 & 69 & Req-Req & \cite{wang2021analyzing, schlutter2020trace, li2024mltracer}\\
Giraph & 811& 688 & 706 & Issue-Commit & \cite{lan2023btlink, zhu2024deep, wang2025mplinker, deng2024promptlink}\\
\multirow{2}{*}{Groovy} & 104 & 100 & 180 & Req-Code & \cite{gao2022using, gao2022propagating, dong2022semi}\\
& 6328 & 6429 & 8851 & Issue-Commit & \cite{dai2023constructing, zhang2023ealink, deng2024mtlink, mazrae2021automated}\\
Groovy & 2940 & 3069 & 987 & Isuue-Issue & \cite{dai2023constructing}\\
grpc & 2957 & 2689 & 3053 & Issue-Commit & \cite{majidzadeh2024multi}\\
\multirow{2}{*}{gson} & 138 & 135 & 143 & Issue-Commit& \cite{majidzadeh2024multi}\\
& 29 & 50 & 55 & Test-Code & \cite{sun2024method, white2022tctracer}\\
guava & 503 & 467 & 548 & Issue-Commit & \cite{majidzadeh2024multi}\\
\multirow{2}{*}{hbase} & 18140 & 14158 & 13379 & Issue-Commit & \cite{dai2023constructing}\\
& 14476 & 15154 & 7694 & Issue-Issue & \cite{dai2023constructing}\\
hibernate & 10527 & 7216 & 6942 & Issue-Commit & \cite{dai2023constructing}\\
& 7513 & 7519 & 3110 & Issue-Issue & \cite{dai2023constructing}\\
\multirow{2}{*}{hive} & 16629 & 11005 & 11058 & Issue-Commit & \cite{dai2023constructing}\\
& 14953 & 15332 & 9110 & Issue-Issue & \cite{dai2023constructing}\\
ignite & 8760 & 8409 & 9997 & Issue-Commit & \cite{zhang2023ealink, deng2024mtlink, mazrae2021automated}\\
iluwater & 340 & 431 & 495 & Issue-Commit & \cite{majidzadeh2024multi}\\
Infinispan & 232 & 4105 & 1117 & Req-Code & \cite{gao2022propagating, gao2022using, dong2022semi}\\
InfusionPump & 21 & 126 & 132 & Req-Req & \cite{schlutter2020trace}\\
Isis & 2135 & 10360 & 10475 & Issue-Commit & \cite{zhu2024deep, lan2023btlink, zhang2023ealink, wang2025mplinker, deng2024promptlink, mazrae2021automated}\\
iTrust & 34 & 137 & 256 & Req-Code & \cite{moran2020improving, wang2021analyzing, fuchss2025lissa, peng2023enhancing, gao2022propagating, zhang2021recovering, gao2022using, zou2024xwcode, li2024mltracer, shen2021supporting, wang2023df4rt, hammoudi2021tracerefiner, hey2021improving, zou2024enhancing}\\
Jbosstools & 6242 & 6905 & 3590 & Issue-Issue & \cite{tian2023cross}\\
jfreechart & 28& 44&44&Test-Code & \cite{sun2024method, white2022tctracer}\\
json-iterator & 70 & 83 & 88 & Issue-Commit & \cite{majidzadeh2024multi}\\
\multirow{2}{*}{Kafka} & 5332 & 4153 & 2893 & Issue-Commit & \cite{dai2023constructing}\\
& 3794 & 4071 & 1699 & Issue-Issue & \cite{dai2023constructing}\\
keras & 533 & 552 & 552 & Issue-Commit & \cite{wang2025mplinker, deng2024promptlink, lin2021traceability, zhu2022enhancing, zhu2024deep,lan2023btlink}\\
\multirow{2}{*}{keycloak} & 5339 & 8590 & 5443 & Issue-Commit & \cite{dai2023constructing}\\ 
& 3302 & 3440 & 1408 & Issue-Issue & \cite{dai2023constructing}\\
KOGITO & 294 & 339 & 90 & Issue-Issue & \cite{tian2023cross}\\
konlpy & 32 & 33 & 33 & Issue-Commit & \cite{white2022tctracer, sun2024aviate}\\
kubernetes-client & 380 & 383 & 389 & Issue-Commit & \cite{majidzadeh2024multi}\\
\multirow{2}{*}{LibEST} & 52 & 14 & 204 & Req-Code & \cite{moran2020improving, gao2024triad}\\
& 52 & 21 & 357 & Req-Test & \cite{moran2020improving}\\
\multirow{3}{*}{maven} & 36 & 786 & 152 & Req-Code & \cite{gao2022propagating, gao2022using, dong2022semi} \\
& 4300 & 6613 & 2596 & Issue-Commit & \cite{dai2023constructing}\\
& 3945 & 3885 & 1971 & Issue-Issue & \cite{dai2023constructing}\\
log4net & 207 & 249 & 248 & Issue-Commit & \cite{wang2025mplinker, deng2024promptlink, zhu2024deep}\\
Mendix & 19 & 221 & 234 & Issue-Commit & \cite{van2023effectiveness}\\
Modis & 19 & 49 & 41 & Req-Req & \cite{du2020automatic}\\
mybatis3 & 350 & 420 & 440 & Issue-Commit & \cite{majidzadeh2024multi}\\
nacos & 132 & 161 & 161 & Issue-Commit & \cite{sun2024aviate, liu2020traceability}\\
NCNN & 96 & 99 & 99 & Issue-Commit & \cite{sun2024aviate, liu2020traceability}\\
netbeans & 1468 & 1269 & 1369 & Issue-Commit & \cite{zhang2023ealink, deng2024mtlink, mazrae2021automated}\\
netty-socketio & 94 & 105 &  112 & Issue-Commit & \cite{majidzadeh2024multi}\\
Nutch & 1755 & 1589 & 1792 & Issue-Commit & \cite{zhu2024deep, deng2024mtlink, deng2024promptlink, wang2025mplinker}\\
OODT & 713 & 940 & 1254 & Issue-Commit & \cite{deng2024mtlink, lan2023btlink, deng2024promptlink, wang2025mplinker}\\
pegasus & 160 & 160 & 160 & Issue-Commit & \cite{sun2024aviate, liu2020traceability}\\
pgcli & 481 & 529 & 529 & Issue-Commit & \cite{lin2021traceability, zhu2022enhancing, zhu2024deep}\\
Pig & 87 & 1316 & 547 & Req-Code & \cite{gao2022propagating, gao2022using, dong2022semi, wang2023df4rt}\\
projectquay & 40 & 40 & 10 & Issue-Issue & \cite{tian2023cross}\\
QMUI-Android & 70 & 70 & 71 & Issue-Commit & \cite{sun2024aviate, liu2020traceability}\\
QMUI-IOS & 32  & 35 & 35 & Issue-Commit & \cite{sun2024aviate, liu2020traceability}\\
quarkus & 5599 & 5625 & 5707 & Issue-Commit & \cite{majidzadeh2024multi}\\
\multirow{2}{*}{railo} & 2315 & 1149 & 989 & Issue-Commit & \cite{dai2023constructing}\\
& 171 & 174 & 45 & Issue-Issue & \cite{dai2023constructing}\\
rax & 532 & 571 & 571 & Issue-Commit & \cite{liu2020traceability, sun2024aviate}\\
realm & 1187 & 1225 & 1267 & Issue-Commit & \cite{majidzadeh2024multi}\\
rxjava & 1765 & 1748 & 1816 & Issue-Commit & \cite{majidzadeh2024multi}\\
san & 186 & 275 & 275 & Issue-Commit& \cite{liu2020traceability, sun2024aviate}\\
Scrapy & 903& 1033 &1034& Issue-Commit & \cite{zhu2024deep}\\
Seam2 & 189 & 1156 & 463 & Req-Code & \cite{gao2022propagating, gao2022using}\\
skylot & 370 & 361 & 395 & Issue-Commit & \cite{majidzadeh2024multi}\\
smos & 67 & 99 & 1027 & Req-Code & \cite{moran2020improving, fuchss2025lissa, peng2023enhancing, li2020combining, du2020automatic, chen2021self, zou2024xwcode, hey2021improving, zou2024enhancing}\\
\multirow{2}{*}{spark} & 20673 & 19549& 14306 & Issue-Commit & \cite{dai2023constructing}\\
& 16105 & 16729 & 8070 & Issue-Issue & \cite{dai2023constructing}\\
square & 252 & 246 & 254 & Issue-Commit & \cite{majidzadeh2024multi}\\
\multirow{2}{*}{switchyard}& 2845 & 2012 & 2428&Issue-Commit&\cite{dai2023constructing}\\
& 2019 & 1974 & 882 & Issue-Issue & \cite{dai2023constructing}\\
\multirow{2}{*}{teiid}& 4826 & 6090 & 5858 & Issue-Issue & \cite{dai2023constructing}\\
& 2058 & 2115 & 723 & Commit-Issue & \cite{dai2023constructing}\\
Tez & 2645 & 2589 & 2785 & Issue-Commit & \cite{zhu2024deep}\\
Tika & 2188 & 3288 & 3163 & Issue-Commit & \cite{zhu2024deep, lan2023btlink}\\
WARC & 21 & 43 & 58 & Req-RReq & \cite{wang2020automated, gao2024triad, schlutter2020trace, rodriguez2021leveraging}\\
weui & 154 & 159 & 159 & Issue-Commit & \cite{liu2020traceability, sun2024aviate}\\
xLua & 52 & 52 & 52 & Issue-Commit & \cite{liu2020traceability, sun2024aviate}\\
\multirow{2}{*}{zookeeper}& 2540 & 1588 & 1423 &Issue-Commit& \cite{dai2023constructing}\\
& 2074 & 2095 & 966 &Issue-Issue& \cite{dai2023constructing}\\

\end{longtable}
\FloatBarrier
{\footnotesize
\noindent \textbf{Note}: Due to their non-open-source nature, we were unable to obtain the following projects: Pine \cite{wang2021analyzing}, MediaStore \cite{keim2024recovering,fuchss2025lissa}, TeaStore \cite{keim2024recovering, fuchss2025lissa}, TEAMMATES \cite{keim2024recovering, fuchss2025lissa}, BigBlueButton \cite{keim2024recovering, fuchss2025lissa}, JebRef \cite{keim2024recovering, fuchss2025lissa}, Dronology (RE) \cite{fuchss2025lissa}, Dronology (DD) \cite{fuchss2025lissa}, Jenkins-core \cite{sun2024method}, Dubbo-rpc \cite{sun2024method}, GanttProject (Requirements-Code) \cite{gao2022propagating, gao2022using}, CAF \cite{marcen2020traceability}, Kromaia 
 \cite{veron2024improving}, Chess 
 \cite{hammoudi2021tracerefiner}, JHotDraw 
 \cite{hammoudi2021tracerefiner}, React \cite{bai2024improving}, Vue \cite{bai2024improving}, and various Android projects \cite{alshara2023ml}. One study did not provide a ground truth dataset \cite{yasa2025evaluating}, three studies used their own proprietary projects as case studies \cite{hassine2024llm, khlif2022complete, yoo2024building}.
}

\section{Results of Co-occurrence Word Ratios for Different Project Types}
The results of RQ2 on co-occurrence word ratios for different project types are provided in Table \ref{table app B}:

\begin{longtable}{@{}p{2.3cm}p{2.2cm}p{1.5cm}p{1.5cm}p{1.5cm}p{1.5cm}@{}}
\caption{Co-occurrence Word Ratios for Different Project Types (where ``TSWCR'' denotes True Set Word Co-occurrence Ratio, ``NTSWCR'' denotes Non-True Set Word Co-occurrence Ratio and ``Req'' denotes ``Requirements'')} \label{table app B} \\

\toprule
Source/Target artifacts types & Projects & TSWCR& NTSWCR  & Difference ratio & p-value \\

\midrule
\endfirsthead

\multicolumn{6}{c}%
{{\bfseries \tablename\ \thetable{} -- continued from previous page}} \\
\toprule
Source/Target artifacts types & Projects & TSWCR& NTSWCR  & Difference ratio & p-value \\
\midrule
\endhead

\midrule \multicolumn{5}{r}{{Continued on next page}} \\
\endfoot

\bottomrule
\endlastfoot

\multirow{18}{*}{Req-Code} &Albergate& 16.02\% & 12.61\% & +27.04\%&$<0.01$ \\
&CCHIT& 35.31\% & 20.91\% &+68.87\% & $<0.01$\\
&Derby & 32.71\% & 15.30\% & +113.79\% &$<0.01$ \\
&Dronology& 19.52\% & 7.87\% &  +148.03\% & $<0.01$\\
& Drools & 19.04\% & 8.24\% & +131.07\%&$<0.01$ \\
&eAnci& 23.35\% & 21.29\% & +9.68 \%&$<0.01$ \\
&EasyClinic&18.47\% & 12.37\% & +49.31\% & $<0.01$ \\
&EBT&19.70\% & 7.10\% & +177.46\% & $<0.01$\\
&eTour& 21.11\% & 11.49\% & +83.72\%&$<0.01$ \\
&Groovy&34.12\% & 19.01 \% & +79.48\% & $<0.01$\\
&Infinispan  & 20.77\% & 8.44\% & +146.09\%&$<0.01$ \\
&iTrust & 29.02\% & 17.02\% & +70.51 \%&$<0.01$ \\
&LibEST&33.54\% & 20.16\% & +66.37\% & $<0.01$\\
&maven& 25.64\% & 10.68\% & +140.07\%&$<0.01$ \\
&Pig  & 25.38\% & 13.37\% & +89.83\%&$<0.01$ \\
&Seam2 & 24.58\% & 8.26\% & +197.58\% &$<0.01$ \\
&smos& 2.07\% & 2.14\% & -3.27\% &0.9009\\
&\textbf{Avg}&\textbf{23.55\%}&\textbf{12.72\%}&\textbf{+93.86\%}&-\\
\midrule
\multirow{9}{*}{Issue-Issue}& archiva & 25.80\% & 7.64\% & +237.70\% & $<0.01$ \\
&automation\_club& 36.29\% & 9.79\% & +270.68\% &$<0.01$\\
&Beam & 28.73\% & 5.48\% & +424.27\% & $<0.01$\\
&Cassandra& 28.28\% & 7.40\% & +282.16\% & $<0.01$\\
&Derby & 31.12\% & 10.06\% & +209.34\% & $<0.01$\\
&Drools & 25.30\% & 7.32\% & +245.63\% & $<0.01$\\
&errai& 26.23\% & 8.71\% & +201.15\% & $<0.01$\\
&feedHenry& 32.39\% & 4.02\% & +705.72\% & $<0.01$\\
& flink & 31.56\% & 6.55\% & +381.83\% & $<0.01$\\
\multirow{16}{*}{Issue-Issue}&Groovy & 32.31\% & 12.73\% & +153.81\% & $<0.01$\\
&hbase & 25.41\% & 7.30\% & +248.08\% & $<0.01$\\
&hibernate& 32.52\% & 10.17\% & +219.76\% & $<0.01$\\
&hive& 25.51\% & 6.88\% & +270.78\% & $<0.01$\\
&Jbosstools& 49.19\% & 7.39\% & +565.63\% & $<0.01$\\
&Kafka& 27.74\% & 8.76\% & +216.67\% & $<0.01$\\
&keycloak& 31.92\% & 6.31\% & +405.83\% & $<0.01$\\
&KOGITO& 26.89\% & 5.20\% & +417.12\% & $<0.01$\\
&maven & 28.43\% & 10.98\% & +158.92\% & $<0.01$\\
&projectquay & 27.84\% & 6.18\% & +350.49\% & $<0.01$\\
&railo& 38.63\% & 7.73\% & +399.74\% & $<0.01$\\
&spark& 31.29\% & 6.76\% & +362.87\% & $<0.01$\\
&switchyard& 26.51\% & 7.36\% & +260.19\% & $<0.01$\\
&teiid& 21.46\% & 6.81\% & +215.12\% & $<0.01$\\
&zookeeper & 25.66\% &7.93\% &+223.58\%&$<0.01$\\
&\textbf{Avg}&\textbf{29.88\%}&\textbf{7.73\%}&\textbf{+309.46\%}&-\\
\midrule
\multirow{2}{*}{Design-Code}& Dronology  & 26.42\% & 7.37\% & +258.48\% & $<0.01$\\
&\textbf{Avg} & \textbf{26.42\%} & \textbf{7.37\%} & \textbf{+258.48\%} & $<0.01$\\
\midrule
\multirow{5}{*}{Req-Req} & Gannt & 32.43\% & 15.22\% & +113.07\% &$<0.01$\\
& InfusionPump & 21.47\% & 7.50\% & +186.27\% & $<0.01$\\
& Modis &  26.48\% & 14.49\% & +82.76\% & $<0.01$\\
& WARC &  36.12\% & 16.15\% & +123.65\% & $<0.01$\\
& \textbf{Avg}&\textbf{29.13\%} & \textbf{13.34\%} & \textbf{+126.44\%} & 
$<0.01$\\
\midrule
\multirow{3}{*}{Req-Design}&CM1-NASA & 27.96\% & 9.27\% & +201.62\% &$<0.01$\\
&Dronology &  61.87\% & 25.77\% & +140.09\% & $<0.01$\\
& \textbf{Avg} & \textbf{44.92\%} & \textbf{17.52\%} & \textbf{+170.86\%} & -\\
\midrule
\multirow{6}{*}{Test-Code}& commons-io&64.05\% & 12.22\% & +424.14\% & $<0.01$\\
& Commons\_Lang & 72.46\% & 9.53\% & +660.33\% & $<0.01$\\
&EasyClinic  & 25.99\% & 17.63\% & +47.42\% & $<0.01$\\
& gson &  34.55\% & 8.07\% & +321.93\% & $<0.01$\\
&jfreechart & 61.75\% & 2.01 \% & +2972.14\% & $<0.01$\\
& \textbf{Avg} & \textbf{51.76\%} & \textbf{9.89\%} & \textbf{+885.19\%} & -\\
\midrule
\multirow{2}{*}{Req-Test}&EasyClinic& 31.85\% & 15.94\% & +99.81\% & $<0.01$\\
& EBT &  45.68\%&18.21\%&+150.85\%& $<0.01$\\
\multirow{2}{*}{Req-Test}& LibEST &  30.81\% & 28.71\% & +7.35\%& $<0.01$\\
& \textbf{Avg} & \textbf{36.11\%} & \textbf{20.95\%} & \textbf{+86.00\%}& -\\
\midrule
\multirow{34}{*}{Issue-Commit}&airflow& 45.22\% & 3.90\% & +1059.49\% & $<0.01$\\
&ant-ivy & 39.04\% & 8.33\% & +368.67\% & $<0.01$\\
&ambari& 47.37\% & 4.84\% & +878.72\% & $<0.01$\\
&archiva & 33.43\% & 5.55\% & +502.34\% & $<0.01$ \\
&arrow & 40.07\% & 5.36\% & +647.57\% & $<0.01$\\
&arthas & 28.27\% & 7.83\% & +261.05\% & $<0.01$\\
&Avro & 40.56\% & 7.44\% & +445.16\% & $<0.01$\\
&awesome-berlin & 20.52\% & 6.52\% & +214.73\% &$<0.01$\\
&Beam & 37.14\% & 4.98\% & +645.78\% & $<0.01$\\
&bk-cmdb & 9.16\% & 3.77\% & +142.97\% & $<0.01$\\
&Buildr & 41.55\% & 8.20\% &  +406.81\%&$<0.01$\\
&bytedeco & 41.37\% & 13.24\% & +212.46\% & $<0.01$\\
&cassandra& 18.03\% & 4.33\% & +316.40\% & $<0.01$\\
&calcite & 35.25\% & 8.45\% & +317.16\% & $<0.01$\\
&canal & 23.45\% & 9.14\% & +156.56\% & $<0.01$\\
&Cica & 24.07\% & 11.36\% & +111.88\% & $<0.01$\\
&Derby & 36.30\% & 7.66\% & +373.89\% & $<0.01$\\
&Drools & 34.42\% & 4.17\% & +725.42\% & $<0.01$\\
&druid & 28.09\% & 10.86\% & +58.66\% & $<0.01$\\
&Emmagee& 20.84\% & 12.99\% & +60.95\% & $<0.01$\\
&errai& 33.78\% & 3.91\% & +763.94\% & $<0.01$\\
&Flask & 10.93\% & 8.84\% &  +23.64\% & $<0.01$\\
&flink& 30.83\% & 5.98\% & +415.52\% & $<0.01$\\
&freemarker& 12.87\% & 3.38\% & +288.77\% & $<0.01$\\
&Giraph& 52.76\% & 5.96\% & +785.23\% & $<0.01$\\
&gson & 65.76\% & 7.76\% & +747.42\% & $<0.01$\\ 
&Groovy & 33.73\% & 12.50\% & +169.84\% & $<0.01$\\
&grpc & 80.10\% & 5.83\% & +1273.93\% & $<0.01$\\
&guava & 44.53\% & 7.08\% & +528.95\% & $<0.01$\\
&hbase & 51.69\% & 5.48\% & +843.25\% & $<0.01$\\
&hibernate& 43.66\% & 5.29\% & +725.33\% & $<0.01$\\
&hive & 46.76\% & 3.96\% & +1080.81\% & $<0.01$\\
&ignite & 32.41\% & 4.85\% & +568.25\% & $<0.01$\\
&iluwater & 42.63\% & 6.79\% & +527.84\% & $<0.01$\\
\multirow{36}{*}{Issue-Commit}&Isis& 18.14\% & 5.46\% & +232.23\% & $<0.01$\\
&json-iterator& 28.71\% & 7.26\% & +295.45\% & $<0.01$\\
&Kafka& 35.54\% & 4.26\% & +734.27\% & $<0.01$\\
&Keras & 36.01\% & 23.17\% & +55.42\% & $<0.01$\\
&keycloak& 36.97\% & 3.62\% & +921.27\% & $<0.01$\\
&konlpy & 17.72\% & 6.12\% & +189.54\% & $<0.01$\\
&kubernate-client& 60.08\% & 4.11\% & +1361.31\% & $<0.01$\\
&log4net &  33.70\% & 8.72\% & +286.47\%&$<0.01$\\
&maven & 27.55\% & 6.96\% & +295.83\% & $<0.01$\\
&Mendix & 15.77\% & 3.28\% & +380.79\% & $<0.01$\\
&mybatis3 & 47.07\% & 5.55\% & +748.11\% & $<0.01$\\
&nacos& 24.24\% & 14.59\% & +66.14\% & $<0.01$\\
&nax& 20.23\% & 7.07\% & +186.14\% & $<0.01$\\
&ncnn & 39.84\% & 7.85\% & +407.52\% & $<0.01$\\
&netbeans& 24.93\% & 9.51\% & +162.15\% & $<0.01$\\
&netty-socketio& 38.15\% & 8.30\% & +359.64\% & $<0.01$\\
&Nutch & 42.43\% & 7.02\% & +504.42\% & $<0.01$\\
&OODT & 34.49\% & 8.15\% & +323.19\% & $<0.01$\\
&pegasus & 51.77\% & 9.08\% & +470.15\% & $<0.01$\\
&pgcli& 11.21\% & 9.37\% & +19.64\% & $<0.01$\\
&QMUI-Android & 27.95\% & 10.84\% & +157.84\% & $<0.01$\\
&QMUI-IOS & 36.61\% & 7.95\% & +360.50\% & $<0.01$\\
&quarkus& 58.60\% & 4.19\% & +1298.57\% &$<0.01$\\
&railo& 1.65\% & 2.92\% & -43.49\% & 1.0000\\
&realm & 60.49\% & 6.79\% & +790.87\% & $<0.01$\\
&RXJava& 57.13\% & 7.66\% & +645.82\% & $<0.01$\\
&san & 21.86\% & 7.96\% & +174.62\% & $<0.01$\\
&Scrapy & 3.55\% & 2.56\% & +38.67\% & $<0.01$\\
&skylot& 41.30\% & 7.03\% & +487.48\% & $<0.01$\\
&spark& 47.76\% & 5.18\% & +822.01\% & $<0.01$\\
&square & 59.11\% & 6.84\% & +764.18\% & $<0.01$\\
&switchyard& 50.19\% & 4.95\% & +913.94\% & $<0.01$\\
&teiid & 22.84\% & 4.14\% & +451.69\% & $<0.01$\\
&Tika & 34.06\% & 7.41\% & +359.65\% & $<0.01$\\
&weui & 15.83\% & 7.87\% & +101.14\% & $<0.01$\\
&xLua&37.45\% & 6.86\% & +445.92\% & $<0.01$\\
\multirow{2}{*}{Issue-Commit}&zookeeper & 44.23\% & 6.23\% & +609.95\% & $<0.01$\\
& \textbf{Avg} & \textbf{35.10\%} & \textbf{7.09\%} & \textbf{+465.19\%} & -\\
\midrule
\multirow{2}{*}{Design-Test}&EasyClinic  & 23.46 \% & 9.72\% & +141.36\% & $<0.01$\\
& \textbf{Avg} & \textbf{23.46\%} & \textbf{9.72\%} & \textbf{+141.36\%} & -\\

\end{longtable}

\end{document}